\documentclass[twocolumn,superscriptaddress,showpacs,amsmath,amssymb,pra,longbibliography]{revtex4-1}

\usepackage[pdftex]{graphicx} 
\usepackage{dcolumn}  
\usepackage{bm}       
\usepackage[usenames,dvipsnames]{color}
\definecolor{URLCOL}{rgb}{0,0.52,0.83} 
\definecolor{LINKCOL}{rgb}{0.05,0.5,0} 
\definecolor{orange}{rgb}{0.6,0.3,0} 
\definecolor{CITECOL}{rgb}{0.25,0,0.48} 
\usepackage{epstopdf}

\definecolor{darkgreen}{rgb}{0.0, 0.5, 0.0}

\usepackage[pdftex,bookmarks,breaklinks,bookmarksopen,bookmarksnumbered,colorlinks,linkcolor=LINKCOL,linktocpage,citecolor=CITECOL,urlcolor=URLCOL,pdfpagemode=UseOutline,pdftex]{hyperref}

\usepackage{booktabs}
\usepackage{comment}
\usepackage{natbib}

\definecolor{TITLECOL}{rgb}{0.1,0.2,0.7} 
\definecolor{SECOL}{rgb}{0.1,0.2,0.7} 
\definecolor{CONTENTSCOL}{rgb}{0.1,0.2,0.7} 
\definecolor{SSECOL}{rgb}{0.25,0,0.48} 
\definecolor{SSSECOL}{rgb}{0.2,0.08,0.53} 
\definecolor{FINCOL}{rgb}{0.01,0.3,0.07} 

\def\ord#1{^{(#1)}} 
\def\gs{^\text{gs}} 
\def\vW{^\text{vW}} 

\def\coloredtitle#1{\title{\textcolor{TITLECOL}{#1}}} 
\def\coloredauthor#1{\author{\textcolor{CITECOL}{#1}}} 

\definecolor{URLCOL}{rgb}{0,0.17,0.43} 
\definecolor{LINKCOL}{rgb}{0.05,0.4,0} 
\definecolor{CITECOL}{rgb}{0.35,0,0.48} 

\def\sss{\scriptscriptstyle\rm}

\def\bea{\begin{eqnarray}}
\def\eea{\end{eqnarray}}
\def\ben{\begin{equation}}
\def\een{\end{equation}}
\def\benu{\begin{enumerate}}
\def\enu{\end{enumerate}}

\def\bei{\begin{itemize}}
\def\eei{\end{itemize}}
\def\beit{\begin{itemize}}
\def\eit{\end{itemize}}
\def\benu{\begin{enumerate}}
\def\enu{\end{enumerate}}

\def\hatT{{\hat T}}
\def\hatV{{\hat V}}
\def\hatH{{\hat H}}
\def\br{{\bf r}}

\def\half{\frac{1}{2}}

\def\PEE{_{{\sss P}}}

\def\s{_{\sss S}}
\def\szero{_{{\sss S},0}}
\def\sone{_{{\sss S},1}}
\def\sj{_{{\sss S},j}}

\def\xc{_{\sss XC}}
\def\HOMO{_{\sss HOMO}}
\def\LUMO{_{\sss LUMO}}

\def\Hxc{_{\sss HXC}}
\def\Hxczero{_{{\sss HXC}}^0}
\def\Hxcone{_{{\sss HXC}}^1}
\def\Hxcj{_{{\sss HXC}}^j}
\def\H{_{\sss H}}
\def\ee{_{\rm ee}}

\def\up{_\uparrow}
\def\dn{_\downarrow}

\def\intr{\int d^3r\,}
\def\intrp{\int d^3r'\,}
\def\intrrp{\int d^3r \int d^3r'\,}

\def\n{n}

\def\Eqsref#1{Eqs.~\eqref{#1}}
\def\Eqref#1{Eq.~\eqref{#1}}
\def\Secref#1{Section~\ref{#1}}
\def\Figref#1{Fig.~\ref{#1}}
\def\Ref#1{Ref.~\cite{#1}}

\def\sec#1{\section{\textcolor{SECOL}{#1}}}
\def\ssec#1{\subsection{\textcolor{SSECOL}{#1}}}

\begin{document}

\coloredtitle{
Kohn--Sham calculations with the exact functional
}
\coloredauthor{Lucas O.\ Wagner}
\affiliation{Department of Physics \& Astronomy, University of California, Irvine, CA 92697}
\affiliation{Department of Chemistry, University of California, Irvine, CA 92697}
\affiliation{Department of Theoretical Chemistry and Amsterdam Center
for Multiscale Modeling, FEW, Vrije Universiteit, De Boelelaan 1083, 1081HV Amsterdam,
The Netherlands}
\coloredauthor{Thomas E.~Baker}
\affiliation{Department of Physics \& Astronomy, University of California, Irvine, CA 92697}
\coloredauthor{E.M. Stoudenmire}
\affiliation{Department of Physics \& Astronomy, University of California, Irvine, CA 92697}
\affiliation{Perimeter Institute of Theoretical Physics, Waterloo, Ontario, N2L 2Y5, Canada}
\coloredauthor{Kieron Burke}
\affiliation{Department of Chemistry, University of California, Irvine, CA 92697}
\affiliation{Department of Physics \& Astronomy, University of California, Irvine, CA 92697}
\coloredauthor{Steven R.\ White}
\affiliation{Department of Physics \& Astronomy, University of California, Irvine, CA 92697}
\date{\today}

\begin{abstract}
As a proof of principle, 
self-consistent Kohn--Sham calculations are performed with the {\em exact} exchange-correlation
functional. 
Finding the exact functional for even one trial density requires solving the interacting Schr\"odinger equation many times. 
 The density matrix renormalization 
group method makes this possible for one-dimensional, real-space systems of more than two interacting electrons.
We illustrate and explore the convergence properties of the exact KS scheme
for both weakly and strongly correlated systems.  We also explore the spin-dependent
generalization and densities for which the functional is ill defined.
\vskip 0.5cm 
{\centering
\includegraphics[width=0.7\textwidth]{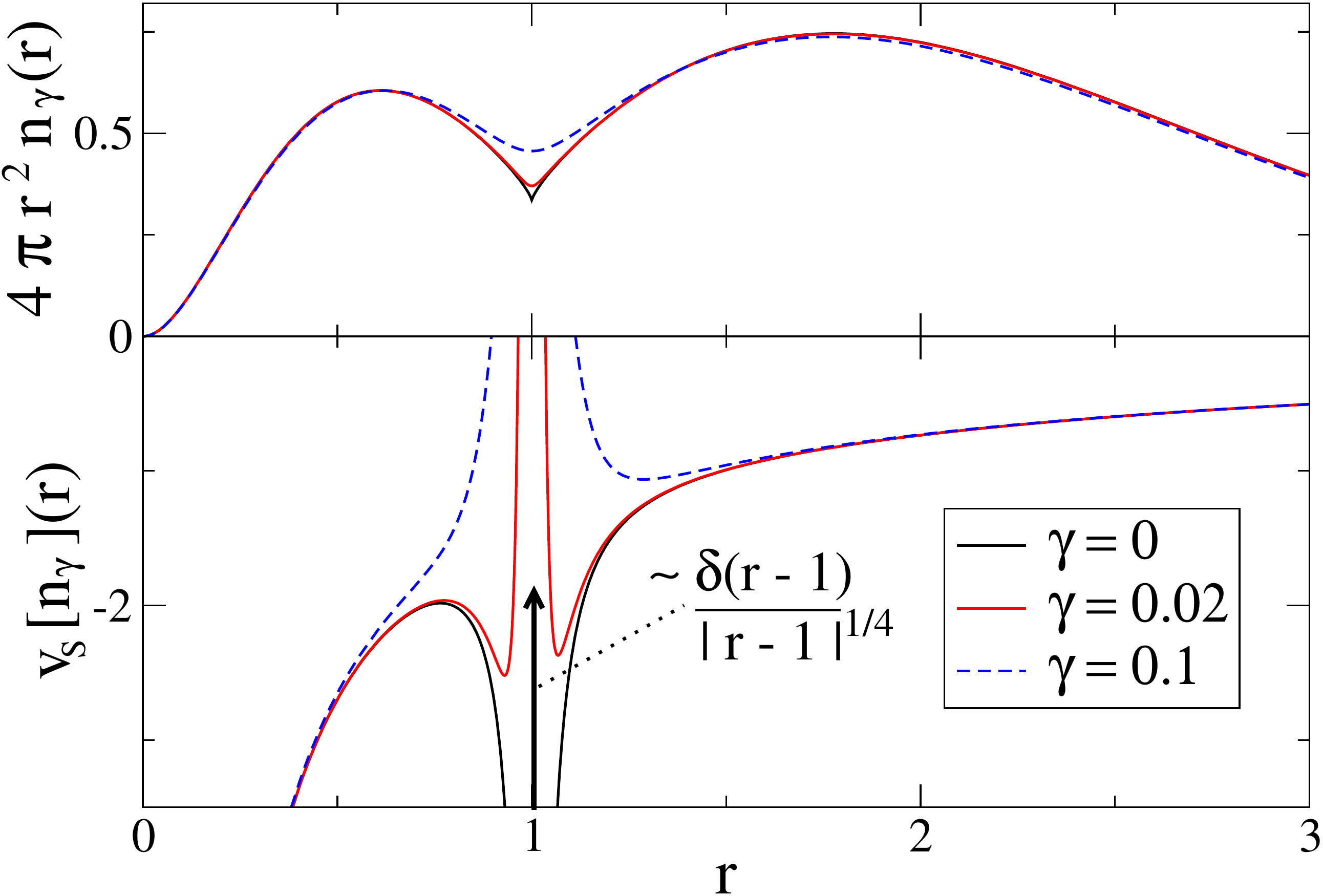}
\vskip 0.01cm
Example of a {\em reasonable} density (with finite kinetic energy) whose Kohn-Sham potential
exists on any grid, but is unphysical in the continuum limit.
}
\end{abstract}

\pacs{71.15.Mb, 
31.15.E-, 
05.10.Cc 
}

\maketitle
\tableofcontents

\sec{Introduction}

Four score and seven years ago, our physics forebears \cite{T27,F28} brought into
this world a new theory, conceived in simplicity, and dedicated
to the proposition that although all particles are waves \cite{S26}, 
their density can be simply calculated \cite{T27,F28}.
Now we are engaged in a great electronic structure
debate, testing whether Kohn--Sham theory \cite{KS65},
or any density functional theory \cite{HK64} so conceived and so dedicated, can
endure in the face of strongly correlated systems.
We have come to dedicate a portion of this paper, as a final convergence proof \cite{WSBW13}
 for those who have dedicated their lives to developing the
constrained search \cite{L83} and approximations thereto \cite{PW92,Bb93,PBE96}.
It is altogether fitting and proper that we should prove this.

Kohn--Sham (KS) \cite{KS65} density functional theory (DFT) is now a widely used
electronic structure method, attaining useful accuracy with present approximations \cite{B12}.
The method finds the ground-state energy of a many-electron, interacting
system by solving an effective non-interacting problem.
This non-interacting problem must be solved self-consistently,
because its potential (the KS potential) is  a 
functional of the electron density.  
The most vital piece of this KS potential is derived
from the mysterious exchange-correlation functional,
which can be computed exactly with great cost \cite{L79,CCD09}. 
This exact functional provides the formal foundations of KS-DFT for
all electronic systems (with some caveats) \cite{L83}. 
However, the utility of KS-DFT derives from simple 
and computationally efficient
approximations to the exchange-correlation (XC) energy \cite{PW92,Bb93,PBE96} which can
be surprisingly reliable and usefully accurate for broad classes of systems,
yet fail badly for others.

Traditionally, study of the exact XC energy functional focused on finding
general exact properties that can either be built into approximations,
or used to understand their failures \cite{LP85,PBE96,PK03,PRTS05}. 
In studying the exact theory, we learn what is and is not
reproduced by the exact functional; e.g.\ that the HOMO-LUMO gap of the KS system is not equal to the 
fundamental (charge) gap of the system \cite{PL83,SS83}. 
As computational power and algorithms evolved, it also became possible
to take a highly accurate solution of the Schr\"odinger equation, extract
the ground-state density, and find the exact KS potential for the system of
interest, notably for few electron systems \cite{AP84,UG94,LB94,FUT94,ZMP94,GLB95,FGU96,LGB96,GB96,WSBW12}. 
These {\em inversions} are often quite demanding, since all quantities must be
sufficiently accurate to extract the small differences in energies and potentials
that form the various components of exchange and correlation.

But even such heroic efforts do {\em not} produce a way of solving the
KS equations with the exact XC functional.   This is because, in an actual
KS calculation, the XC functional is needed not just for the ground-state
density of the system to be solved, but for a sequence of trial densities
that ultimately converges to the solution for that problem.   To
find the XC functional for some trial density, one must solve the Schr\"odinger
equation for the potential for which that density is the ground state, both
for interacting and non-interacting electrons.  Worse still, these potentials are a priori
unknown.  Advancing just one step in the
KS calculations thus requires solving many interacting electronic
problems in order to {\em find} the right potential that yields the trial density.
We call this an interacting inversion, and previous examples have been
limited to 2 electrons \cite{TGK08,V08,CCD09}.

In this paper, we detail how to find the exact XC functional for realistic models of 
electrons in one dimension.  By realistic, we mean models whose 
properties mimic those of real systems, and whose treatment with
approximate density functionals yields results similar to those for real systems \cite{WSBW12}.
We use the density matrix renormalization group \cite{White:1992,White:1993a,Schollwoeck:2005}
 to solve the Schr\"odinger equation, because of its tremendous efficiency
for one-dimensional (1d) systems.
In \Ref{WSBW13}, we used this capability to explore the convergence of a simple
algorithm for the KS scheme, ultimately proving that, no matter how
strongly correlated, convergence can always be achieved in a finite number
of iterations.  Various approximate functionals have their own convergence
proofs \cite{CL00,C00}, but here we detail exactly how the exact calculations are done,
and test further properties of the exact functional.

\sec{Background}\label{s:back}

Typical solid state and quantum chemistry investigations into electronic
structure begin with the non-relativistic continuum Hamiltonian in the Born--Oppenheimer
approximation, 
\bea
\hatH &\equiv& \hatT + \hatV + \hatV\ee \label{eqn:H}\\
&\equiv& \sum_{i=1}^{N}\left( -\half  \nabla_i^2  + v(\br_i) \right)
				 + \half \sum_{i\neq j}^{N} \dfrac{1}{|\br_i - \br_j|}, \nonumber
\eea
which describes the quantum behavior of $N$ electrons in an external potential $v(\br)$
determined by the (classical) nuclei via the operators:
the electron kinetic energy $\hatT$, their potential energy $\hatV$,
and the electron-electron interaction $\hatV\ee$.  The eigenstates $\Psi_j$ 
and eigenvalues $E_j$ (the energies) 
of the Hamiltonian  $\hatH$
determine all the properties of the system.

Despite \Eqref{eqn:H} being the key to everyday electronic structure,
an accurate solution for even the ground-state energy $E$ and wavefunction
 $\Psi$ is not presently tractable for large molecules.  
This problem continues to inspire the development of new approximations
and methods to solve the many-body problem.  Some methods---such 
as Hartree--Fock theory \cite{SO82}, quantum Monte Carlo \cite{NU99},  
and coupled cluster \cite{SO82}---attempt 
to approximate, sample, or construct the wavefunction.
Density functional theory, on the other hand, approaches the many-body problem
quite differently.

While $\Psi$
allows one to characterize the system completely,
the much simpler ground-state electron density $n(\br)$ was proven by Hohenberg and Kohn (HK)
to also determine all the properties of the system \cite{HK64}. 
  Their theorem allows us to formally work
with the density as the basic variable instead of the wavefunction \cite{L83}. 
The keystone of this far-reaching proof is the one-to-one correspondence
between the ground-state density $n(\br)$ and the potential $v(\br)$ of
a system, which characterizes the system completely.  
This one-to-one mapping  will be explored in greater detail 
in \Secref{s:inv}, since it is crucial for calculate
the exact functional.

As an important mathematical aside, the potential corresponding
to a given density is unique if it exists, but there are
 some densities $n(\br)$ which are not {\em ensemble $v$-representable,} i.e.\
 not the ground states of any potential $v(\br)$ \cite{EE83}. 
We explore this complication later, in \Secref{s:nonvrep}.

A simple corollary of the HK theorem is that the ground-state energy 
of a system can be determined by minimizing over trial electron densities \cite{HK64}
\bea
E_v &=& \min_n  E_v[n] \label{eqn:E0} \\
E_v[n] &\equiv&  F[ n] + \intr   n(\br)\, v(\br),  \label{eqn:Evn} 
\eea
where $F[n]$ accounts for the electronic kinetic energy and electron-electron 
repulsion energy, and is universal, i.e., independent of the external potential $v(\br)$.
When degeneracy is not an issue \cite{L82}, 
the functional $F[n]$ can  be found by minimizing the expectation value of
$\hatT + \hatV\ee$ over all properly antisymmetric wavefunctions $\Psi$ that
yield the density $n(\br)$ \cite{L79,L83}:
\ben
F[ n] = \min_{\Psi \to n} \langle \Psi |\{ \hatT + \hatV\ee \} | \Psi \rangle, \label{eqn:F}
\een
and the minimizing $\Psi$ is denoted $\Psi[n]$.  This is the {\em pure-state} formulation
of DFT.  The generalization for degenerate systems involves replacing the expectation value in Eq.~\eqref{eqn:F} 
with a trace over the ground-state ensemble $\Gamma$ \cite{L83}.
The only known way to exactly calculate the functional
thus implicitly requires use of a wavefunction (or a density matrix for degenerate 
systems).

We now turn to the formulation of the most popular of DFT implementations,
Kohn--Sham DFT \cite{KS65}.  Kohn--Sham theory 
creates a doppelg\"anger of the interacting system:  a set of
non-interacting electrons with the same density.
This non-interacting system, the KS system, is characterized by its potential,
$v\s[n](\br)$, defined implicitly so
that a system of $N$ non-interacting electrons in this potential
has density $n(\br)$.
This means that after solving the non-interacting Schr\"odinger (i.e.\ KS) equation and obtaining
the KS orbitals $\phi_j(\br)$ (in Hartree units):
\ben
\left\{ -\half \nabla^2 + v\s[n](\br) \right\} \phi_j(\br) = \epsilon_j\, \phi_j(\br). \label{eqn:KSeqn}
\een
One finds the density $n(\br)$ by occupying the $N/2$ lowest-energy orbitals,
\ben
n(\br) = 2 \sum_{j=1}^{N/2} |\phi_j(\br)|^2 \label{eqn:dens}
\een
(where  for simplicity we assume that
the system is spin-unpolarized). 
Obtaining the KS potential $v\s[n](\br)$ for a density $n(\br)$ is an
inverse problem, on a firm foundation through the HK theorem applied to non-interacting systems.
(Some densities, however, will prove to be non-$v$-representable \cite{Leeuwen}, so the potential $v\s[n](\br)$
is unique, up to a constant, but only if it exists.)
Many algorithms to invert a density to find its KS potential have been suggested \cite{G92,WP93,LB94,ZMP94,LGB96,SGB97,PNW03}; 
ours will be described in \Secref{s:inv}.

As a descendent of DFT, Kohn--Sham DFT  determines the energy of a system 
by knowledge of the density alone.  
Within the KS framework, the universal functional $F[n]$ is written as
\ben
F[n] = T\s[n] + U[n] + E\xc[n] \label{eqn:FKS}
\een
where $T\s[n]$ is the kinetic energy of the KS orbitals:
\ben
T\s[n] \equiv -\sum_{j=1}^{N/2}  \intr \phi_j^*(\br) \nabla^2 \phi_j(\br) ,\label{eqn:Ts}
\een
$U[n]$ is the Hartree energy:
\ben
U[n] \equiv \half \intr \!\intrp \dfrac{n(\br)\, n(\br')}{|\br-\br'|},\label{eqn:U}
\een
and $E\xc[n]$ is the exchange-correlation (XC) energy, defined by \Eqref{eqn:FKS}.
Very successful (albeit crude) approximations to $E\xc[n]$ have been developed \cite{PW92,Bb93,PBE96}, 
 which make KS theory a standard and practical approach to electronic structure.
Our work focuses on the exact $E\xc[n]$, with a few comparisons to the simplest density functional approximation,
the local density approximation (LDA) \cite{PW92}.

The KS framework  offers a convenient way to minimize $E_v[n]$ as in 
\Eqref{eqn:E0}, by solving non-interacting systems with an effective potential.
We guess an input density $n_\text{in}\ord{i}(\br)$ and use it to 
calculate a trial KS potential $v\s\ord{i}(\br)$:
\ben
v\s\ord{i}(\br) = v(\br) + v\H[ n\ord{i}_\text{in}](\br) + v\xc[ n\ord{i}_\text{in}](\br), \label{eqn:KSupdate}
\een
where $v\H[n](\br) = \delta U[n] / \delta n(\br)$ is the Hartree potential:
\ben
v\H[n](\br) = \intrp \dfrac{n(\br')}{|\br-\br'|},
\een
and $v\xc[n](\br) = \delta E\xc[n] / \delta n(\br)$ is the XC potential.
The Hartree and XC potentials
together account for two-body interactions \cite{BW12}, and are found by
taking functional derivatives of their parent energy functionals.

\begin{figure}[b]
\unitlength1cm
\begin{picture}(6,7)
\put(0,0){\makebox(6,7){
\includegraphics[width=9cm]{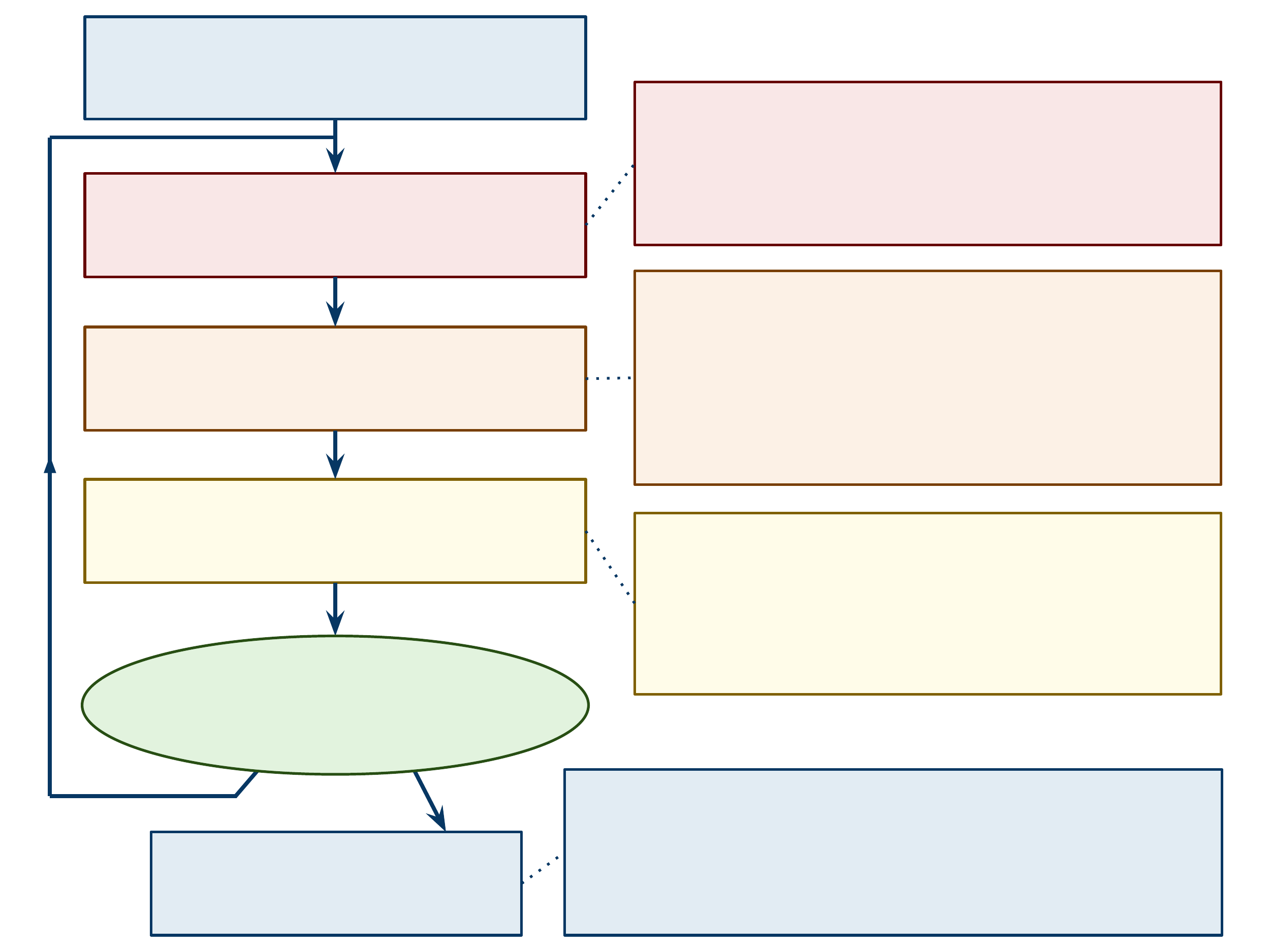}
}}
\put(-1.63,5.87){\makebox(5,1){\centering Guess initial density}}
\put(-1.63,4.76){\makebox(5,1){\centering Create KS potential}}
\put(-1.63,3.66){\makebox(5,1){\centering Solve KS equations}}
\put(-1.63,2.575){\makebox(5,1){\centering Obtain new density}}
\put(-1.63,1.47){\makebox(5,1){\centering Converged?}}
\put(-1.63,0.1){\makebox(5,1){\centering Done}}
\put(0.2,1.54){\footnotesize \it no} 
\put(1.1,1.54){\footnotesize \it yes} 
\put(2.61,5.13){\makebox(5,1.1){\parbox{5cm}{
$\displaystyle v\s(\br):= v(\br) + v\H[n_\text{in}](\br)$\\[2pt]
$\quad\quad\quad\quad+\, v\xc[n_\text{in}](\br)$
}}}
\put(2.61,3.655){\makebox(5,1.1){\parbox{5cm}{ 
$\displaystyle \left \{ -\half \nabla^2 + v\s(\br) \right \}\phi^\text{out}_j(\br) \quad \ $\\
$\quad \quad \quad \quad \quad \quad = \epsilon_j \, \phi^\text{out}_j(\br)  $
}
}}
\put(2.61,2.06){\makebox(5,1.1){\centering 
$\displaystyle n_\text{out}(\br) = 2 \sum_{j=1}^{N/2} |\phi^\text{out}_j(\br)|^2$
}}
\put(2.38,0.3){\makebox(5,1.1){\parbox{5.5cm}{
$E = T\s[n]+\intr n(\br)\,v(\br)$\\[2pt]
$\quad\quad\quad+U[n]+E\xc[n]$
}
}}
\end{picture}
\caption{The KS scheme.}\label{KSscheme}
\end{figure}
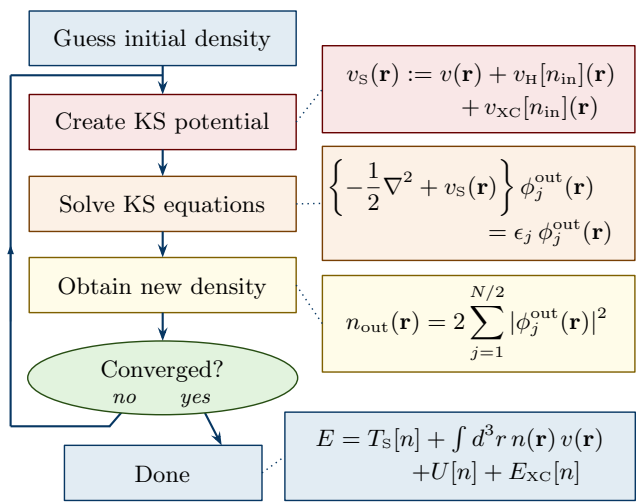

After calculating $v\s\ord{i}(\br)$ for the given input density,
we solve the trial KS system  (i.e.\ \Eqref{eqn:KSeqn} with our trial KS potential)
to obtain an output density $n_\text{out}\ord{i}(\br)$.
If the output density equals the input density, we have achieved self-consistency
and have found a stationary point of $E_v[n]$.  This may be quantified by
calculating a simple criterion for convergence:
\ben
\eta\ord{i} \equiv \dfrac{1}{N^2} \intr \Big(  n\ord{i}_\text{out}(\br) - n\ord{i}_\text{in}(\br) \Big)^2, \label{eqn:tolerance}
\een
declaring the calculation converged when $\eta\ord{i} < \delta$, 
If the calculation has not converged, 
a new guess density $n\ord{i+1}_\text{in}(\br)$, such as $n\ord{i}_\text{out}(\br)$,
 is  plugged into  \Eqref{eqn:KSupdate} for the next iteration, 
and we repeat until converged.  
For the exact XC functional, the converged density is the ground-state density of interacting
electrons in the potential $v(\br)$ \cite{WSBW13}.
This iterative-convergence procedure is known as the KS scheme \cite{DG90}, 
and is illustrated in \Figref{KSscheme}.
The possibility of finding other
stationary points besides the ground-state for the exact
functional will be addressed in  \Secref{s:KS}.

The Kohn--Sham DFT  approach to electronic structure thus
converts the many-body problem into a non-interacting problem
which must be solved self-consistently. 
The exact procedure requires finding the many-body system with 
a given density, with wavefunction $\Psi[n]$, to determine $E\xc[n]$ and $v\xc[n](\br)$,
and thus is as costly as solving the original many-body problem (see \Secref{s:KS}).
However, the KS scheme would be neither useful nor practical at
such a computational cost.  Evaluating $v\xc[n](\br)$ at each iteration of the KS scheme is (usually)
a trivial and inexpensive step with present approximations, since
the functional derivative is known explicitly.

\sec{Inversions}\label{s:inv}

Inverting a density $n(\br)$ to find its KS potential $v\s[n](\br)$,
or to find its external potential $v[n](\br)$ (for real, interacting 
electrons) is not a straightforward task.
In this section we 
discuss how to do this for an arbitrary $v$-representable density.
As a by-product of these inversions, we obtain
the implicitly defined KS orbitals and interacting wavefunction $\Psi[n]$,
which allow us to evaluate the XC potential and energy in \Secref{s:KS}.

{\em Non-interacting inversions} are performed to find the KS potential of exact densities
for a variety of systems \cite{UG94,GLB95,WSBW12}.  The notation we use for the potential corresponding
to the density $n(\br)$ of non-interacting electrons
is $v\s[n](\br)$, which we have already seen in \Eqref{eqn:KSeqn}.
This inversion is a simple matter for one or two electrons with opposite spins,
since the KS equation can be rearranged to obtain:
\ben
v\s[n](\br) = \half \dfrac{\nabla^2 \sqrt{n(\br)} }{\sqrt{n(\br)}} + \epsilon,\quad(N \le 2) \label{eqn:vSW}
\een
where $\epsilon$ is a constant (the only occupied KS eigenvalue).
For more electrons, one can use  an iterative procedure 
 to determine $v\s[n](\br)$.  Initially a potential $v\s\ord{1}(\br)$
is guessed, e.g.\ \Eqref{eqn:vSW}.  Then, starting with $i = 1$:
\benu
\item For the potential $v\s\ord{i}(\br)$,
solve the non-interacting Schr\"odinger equation
for orbitals $\phi_j\ord{i}(\br)$, doubly-occupying to obtain 
the density $n\ord{i}(\br)$. \label{step:guess}

\item If $n\ord{i}(\br)$ is within tolerance of $n(\br)$, we are done,
i.e.\ 
$v\s\ord{i}(\br) = v\s[n](\br)$ and $\phi_j(\br) = \phi\ord{i}_j(\br)$.  Otherwise,
continue.

\item A new potential $v\s\ord{i+1}(\br)$ is chosen, based on how different $n\ord{i}(\br)$
is from $n(\br)$.  Roughly speaking, where $n\ord{i}(\br)$ is too low, 
the new potential $v\s\ord{i+1}(\br)$ is lowered from the old $v\s\ord{i}(\br)$,
and where $n\ord{i}(\br)$ is too high, the new potential is raised. \label{step:update}

\item Increment $i$ and repeat steps \ref{step:guess} to \ref{step:end}. \label{step:end}
\enu
The only difference between different inversion algorithms is how the new
potential is determined in step \ref{step:update}. The problem can
be reduced to finding the root of a nonlinear function of many variables,
which can be treated at various levels of sophistication \cite{NumRec}.
We discuss Broyden's method at the end
of this section.  
With the KS potential $v\s[n](\br)$ and orbitals
$\phi_j(\br)$, we can evaluate functionals such as $T\s[n]$ using \Eqref{eqn:Ts}.

{\em Interacting inversions} are rarely done, since
they are far more expensive than non-interacting inversions,
and require solving the many-body problem many times.
Only two-electron problems have been studied,
in one case
to understand the adiabatic approximation within TDDFT \cite{TGK08,V08}
and in another
to study  the self-interaction error within LDA \cite{CCD09};
though we have recently studied four-electron systems \cite{WSBW13}.
The potential $v[n](\br)$, which corresponds to the
interacting system of electrons with density $n(\br)$,
can be found using the same algorithm as for $v\s[n](\br)$,
though in step \ref{step:guess} we must solve an interacting problem
for the many-body wavefunction $\Psi\ord{i}$ rather than the non-interacting Schr\"odinger equation
for orbitals $\phi\ord{i}_j(\br)$.  At the end of the inversion
we obtain $\Psi[n]$, the wavefunction which minimizes $F[n]$ in \Eqref{eqn:F},
allowing us to compute $F[n]$ for that specific density.

\begin{figure}
\includegraphics[width=\columnwidth]{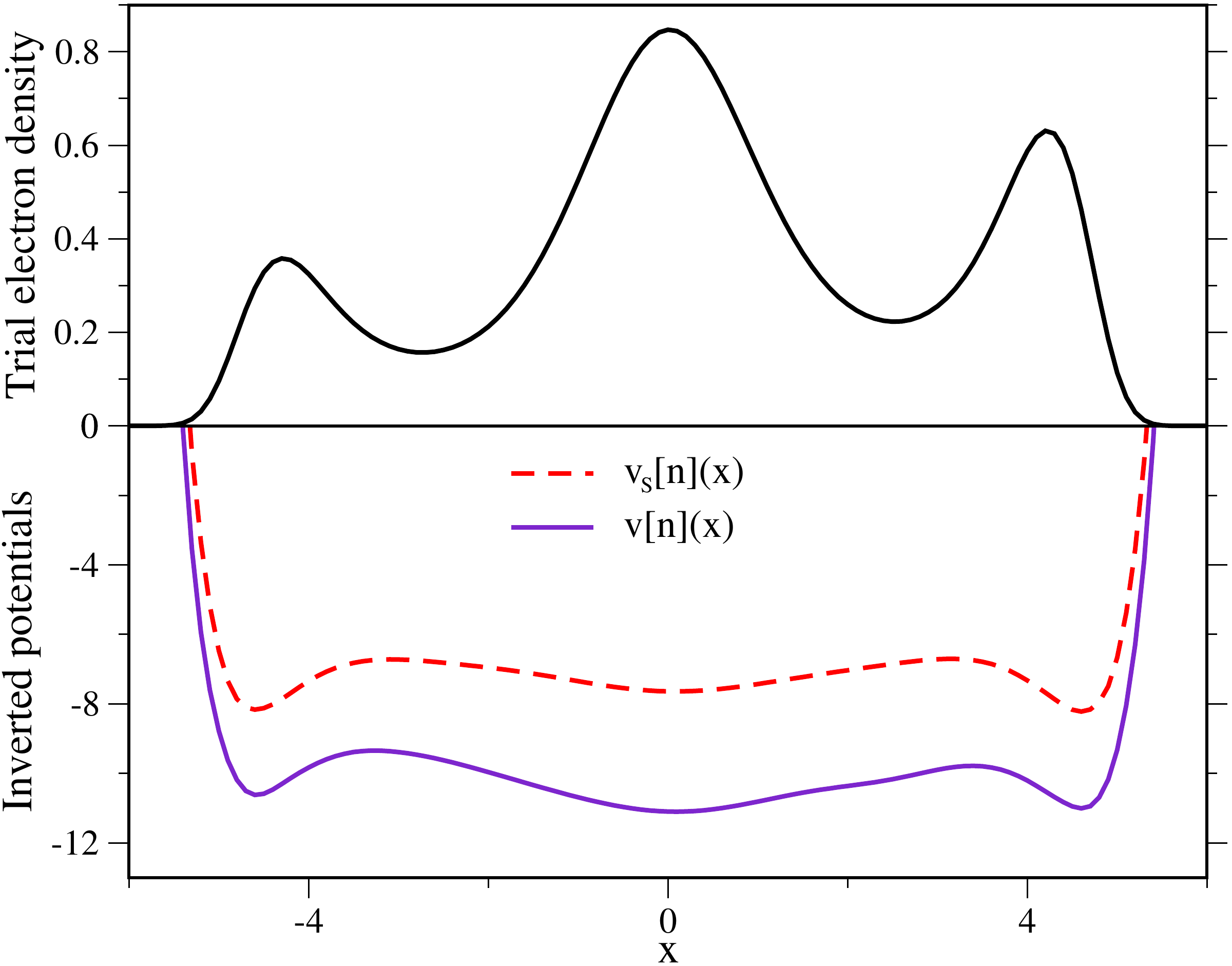}
\caption{
Density inversion of arbitrary 4-electron density for non-interacting and interacting potentials. 
Solving either the interacting Schr\"odinger equation in the potential $v[n](x)$ or solving the 
non-interacting Schr\"odinger equation in the potential $v\s[n](x)$ yields the density in the 
top panel.}
\label{Cactus}
\end{figure}

To illustrate the theory behind KS-DFT,
we solve interacting systems using the density matrix renormalization group (DMRG) \cite{White:1992,White:1993a},
which is the most efficient wavefunction solver in 1d, capable of handling both
strong and weak correlation. We apply DMRG to model 1d continuum systems
by discretizing space into $N_g$ grid points with a small grid spacing $\Delta$  \cite{SWWB12,WSBW12}. 
With this method,  we can invert 1d systems with over 100 electrons \cite{SWWB12}.  
For our model systems we employ a softened Coulomb interaction between electrons \cite{ESJ89,TGK08,HFCV11,SWWB12,WSBW12}:
\ben
v\ee(u) = 1/\sqrt{u^2+1}.\label{eqn:softee}
\een

Figure \ref{Cactus} shows a four-electron example of an interacting inversion
\footnote{The density shown in \Figref{Cactus} is given by 
\mbox{$\tilde{n}(x) = e^{x/15-x^2/2+x^4/20-x^6/750}$}, divided 
by a normalization factor such that it contains four electrons.}.  
For some arbitrary density like this one (meaning a density we would not find
in nature), we want to find the associated KS and interacting potentials.
This is the problem we encounter during the self-consistent calculation of the
KS equations.  Since we ultimately find $\Psi[n]$ at the end of the inversion,
we can evaluate $F[n]$
(given soft-Coulomb interactions); likewise with $\phi_j(\br)$ we can obtain
$T\s[n]$. For the example density of \Figref{Cactus} we find $F[n]=3.07$, $T\s[n]=0.843$,
$U[n] = 3.628$, so $E\xc[n] = -1.397$.
The XC energy is thus calculated using simple energy differences;
and we obtain the XC potential in the same way.  
We further describe these matters in the next section.

To close this section, we describe our recipe for step \ref{step:update} 
of the inversion algorithm.
The idea is to build an approximation 
for the density-density response matrix, $\chi$,
which determines how a small change in the potential will change the density:
\ben
\intrp \chi(\br,\br')\, \delta v(\br') = \delta n(\br).
\een
Restricting our attention to 1d, 
we recast this equation as the matrix
equation $\chi\, \delta v = \delta n$, where $\chi$ is an (unknown) $N_g\times N_g$  matrix,
and $\delta v$, $\delta n$ are vectors with $N_g$ components, where $N_g$
is the number of grid-sites in the system.
A constant change in the potential 
(i.e.\ $\delta v = c_1$) will give zero change in the density ($\delta n = 0$),
and a constant change in the density ($\delta n = c_2$) is impossible, since $N$ is fixed.
Therefore we consider orthonormal basis functions for changes in the potential and density which
integrate to zero, encoded as columns in the matrices $W$ and $M$, respectively \cite{K83}. 
Within this basis,  the density-density response matrix can be approximated by
a smaller matrix, $A$:
\ben
\chi \approx M A\, W^T. \label{chiSVD}
\een
This factorization of the matrix $\chi$ looks very much like (and is inspired by) the 
singular value decomposition (SVD) of $\chi$,
which would give an exact breakdown of $\chi$ into optimal bases $M$ and $W$, with $A$ being diagonal.
We do not
know $\chi$ a priori, but an approximation to $\chi$ (or $A$) can be iteratively
improved using a quasi-Newton method (we use Broyden's method from \Ref{B65}). 
We construct appropriate basis vectors for $M$ and $W$
using orthonormalized differences of trial densities from the target density. 
As $A$ is refined, the bases $M$ and $W$ can be optimized (if desired) by computing the SVD
of $A$, a procedure which is also useful to compute $A^{-1}$, and thus $\chi^{-1}$.
The next trial potential for step \ref{step:update} is determined by:
 $v\ord{i+1} = v\ord{i} + \chi^{-1} (n - n\ord{i})$.
Typically around 20 basis vectors in $M$ and $W$ are required to obtain
a trial density indistinguishable from the target density
on the scale of \Figref{Cactus}.

\sec{Results}\label{s:KS}

\begin{figure}[b]
\unitlength1cm
\begin{picture}(6,7)
\put(0,0){\makebox(6,7){
\includegraphics[width=8.8cm]{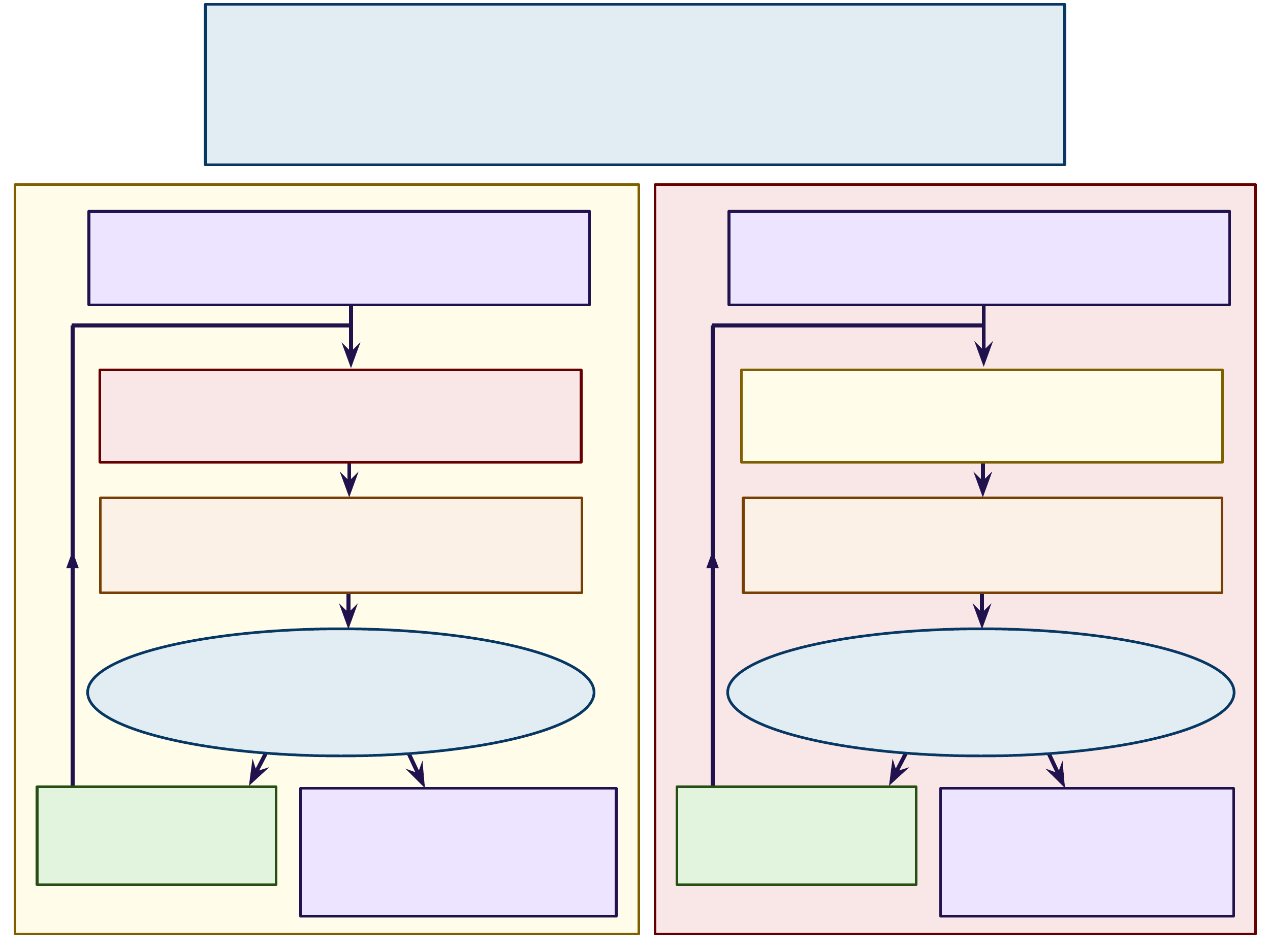}
}}
\put(0,5.96){\makebox(6,1){\centering $E\Hxc[n] = \langle \Psi[n] | \{ \hatT + \hatV\ee \} | \Psi[n] \rangle - T\s[n]$}} 
\put(0,5.45){\makebox(6,1){\centering $v\Hxc[n](\br) = v\s[n](\br) - v[n](\br)$}} 
\put(-1.56,4.50){\makebox(5,1){\centering Guess initial potential}}
\put(1.2,4.38){$\tilde v\s(\br)$} 
\put(-1.55,3.4){\makebox(5,1){\centering Find $\tilde \varphi_j(\br)$ from $\tilde v\s(\br)$}}
\put(-1.55,2.45){\makebox(5,1.1){\centering Get $\tilde n(\br)$ from $\tilde \varphi_j(\br)$}}
\put(-1.55,1.55){\makebox(5,1){\centering $\ \tilde n(\br) = n(\br)$?}}
\put(0.25,1.66){\footnotesize \it no} 
\put(1.15,1.66){\footnotesize \it yes} 
\put(0.8,0.32){\makebox(2,1.1){\parbox{3cm}{
$\displaystyle v\s[n](\br)\! =\!\tilde v\s(\br)$\\
$\phi_j(\br) = \tilde \varphi_j(\br)$
}}}
\put(-1.3,0.43){\makebox(2,1.1){\centering Alter $\tilde v\s(\br)$}}
\put(2.88,4.5){\makebox(5,1){\centering Guess initial potential}}
\put(5.55,4.38){$\tilde v(\br)$} 
\put(2.88,3.4){\makebox(5,1){
\centering Obtain $\tilde \Psi$ from $\tilde v(\br)$ 
}}
\put(2.88,2.45){\makebox(5,1.1){\centering Get $\tilde n(\br)$ from $\tilde \Psi$}}
\put(2.87,1.55){\makebox(5,1){\centering $\ \tilde n(\br) = n(\br)$?}}
\put(4.7,1.66){\footnotesize \it no} 
\put(5.6,1.66){\footnotesize \it yes} 
\put(3.12,0.43){\makebox(2,1.1){\centering Alter $\tilde v(\br)$}}
\put(5.13,0.319){\makebox(2,1.1){\parbox{2cm}{
$\displaystyle v[n](\br)\! =\! \tilde v(\br)$\\
$\Psi[n] = \tilde \Psi$
}}}
\end{picture}
\caption{To determine the $E\Hxc[n]$ and $v\Hxc[n](\br)$:
Our exact calculation  requires a computationally demanding inversion
algorithm  to find the one-body potential $v[n](\br)$ of the
interacting system whose density is $n(\br)$, with KS orbitals 
$\phi(\br)$, in addition to
a non-interacting inversion to find $v\s[n](\br)$.  In case
of degeneracy, mixed-states should be used instead of pure-state
wavefunctions in both non-interacting and interacting inversions \cite{V80,L83}.   
The right hand side differs from the left in that it describes an 
interacting inversion.
}\label{f:EHxc}
\end{figure}
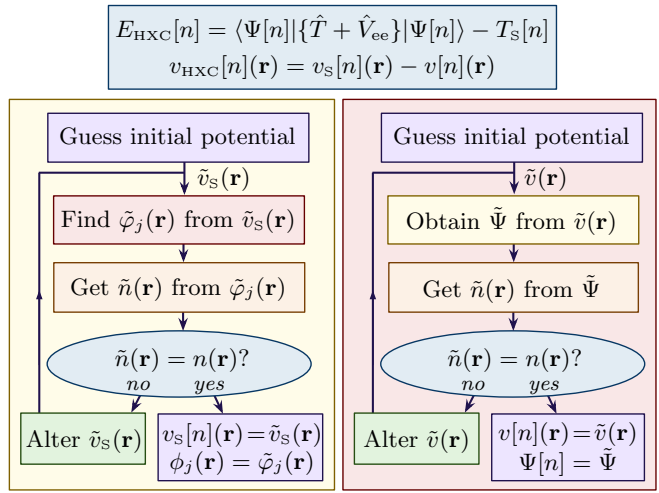

We have now sufficient machinery to calculate
the exact exchange-correlation energy and potential for any trial
density, as encountered in the KS scheme.  
For convenience, we define $E\Hxc[n] \equiv U[n] + E\xc[n]$,
which can be evaluated (using \Eqsref{eqn:F} and \eqref{eqn:FKS}) as:
\ben
E\Hxc[n] = \langle \Psi[n] | \{ \hatT + \hatV\ee \} | \Psi[n] \rangle - T\s[n].\label{eqn:EHxc}
\een
From Section \ref{s:inv}, we know how to obtain $\Psi[n]$ and $T\s[n]$
using inversions.
Therefore the exact $E\xc[n]$ is no obstacle in principle,
but extremely computationally expensive in practice.
Similarly, the HXC potential is:
\ben
v\Hxc[n](\br) = v\s[n](\br) - v[n](\br),\label{eqn:vHxc}
\een
which are available from interacting and non-interacting
inversions.  The construction of the exact functional using inversions is illustrated
in \Figref{f:EHxc}.

To algorithmically implement the KS scheme, we must choose 
our input densities $n_\text{in}\ord{i}(\br)$ for each iteration $i$;
each output density $n_\text{out}\ord{i}(\br)$ is determined by solving
the KS equations \eqref{eqn:KSeqn}.
Although more
sophisticated  algorithms are used in practice \cite{SH73,CL00,DS00,KSC02,TOYJ04,FMM04,YMW07}, 
we choose the simple algorithm given below.
We emphasize that we make no claims as to the efficiency of this particular algorithm.
We expect many other algoritms to be more efficient.  But this simple choice allows a simple
proof of convergence, and provides an initial framework to study convergence rate
questions.

The first input density $n_\text{in}\ord{1}(\br)$ is arbitrarily chosen.
The subsequent input densities are calculated via the linear density mixing 
algorithm,
\ben
n\ord{i+1}_\text{in}(\br) = (1-\lambda)\,n\ord{i}_\text{in}(\br) + \lambda\, n\ord{i}_\text{out}(\br),\label{eqn:densmix}
\een
where $\lambda$ is a parameter between 0 and 1, which aids convergence.
At $\lambda = 1$, no density mixing is performed, and the output density of iteration $i$
is used as the input for iteration $i+1$.  While this might allow for quick
convergence, there is the danger of repeatedly overshooting the ground-state density and not converging.
If this happens, smaller steps must be taken, i.e.\ small $\lambda$ ($\lambda = 0$ not allowed) must be used.
These convergence issues are  discussed more thoroughly in \Secref{s:conv}, 
where we investigate how small this density mixing $\lambda$ needs to be in order
to converge the calculation.

\ssec{Illustration}\label{s:illustration}

In this section we use the exact functional within the KS scheme for a
model one-dimensional continuum system, demonstrating convergence
to the true ground-state density.  We also explain why
 the only stationary point of the exact functional is the true
ground-state density.

In our model one-dimensional system, 
electrons are attracted to the nuclei via the potential \cite{WSBW12}
\ben
v_\text{e-nuc}(x) = -1/\sqrt{x^2+1},
\een
and electrons interact with the corresponding repulsive 
potential as already mentioned via \Eqref{eqn:softee}.

\begin{figure}
\includegraphics[width=\columnwidth]{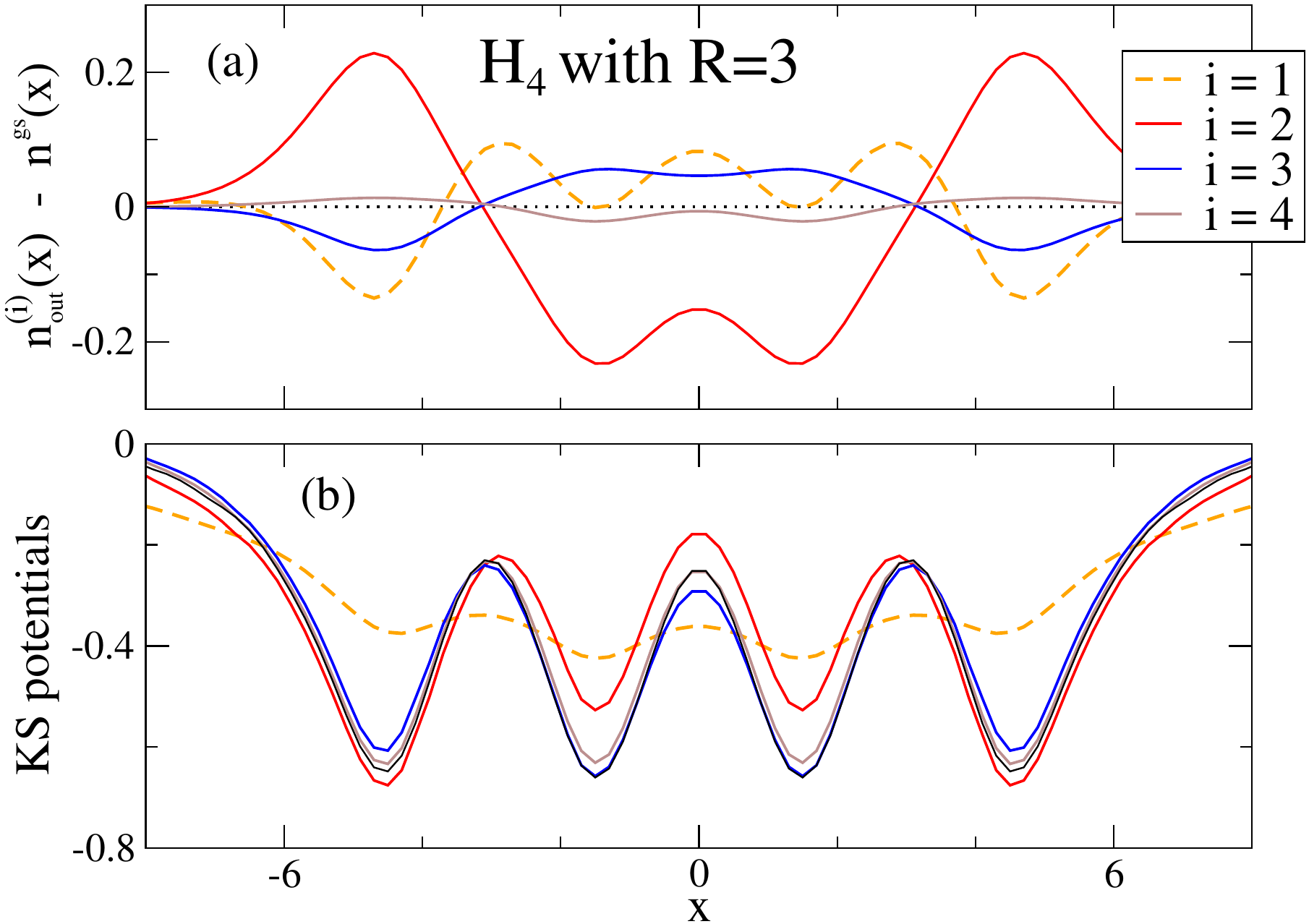}
\caption{KS procedure for a moderately correlated 4-electron system
(four hydrogen atoms separated by an interatomic spacing of $R=3$),
using a fixed $\lambda = 0.3$ and
showing the first few iterations of:
(a) differences in the trial output densities from the ground-state density
(shown in \Figref{H4b3xc})
and (b) trial KS potentials.  Data from \Ref{WSBW13}.
 }
\label{H4b3KS}
\end{figure}

\begin{figure}
\includegraphics[width=\columnwidth]{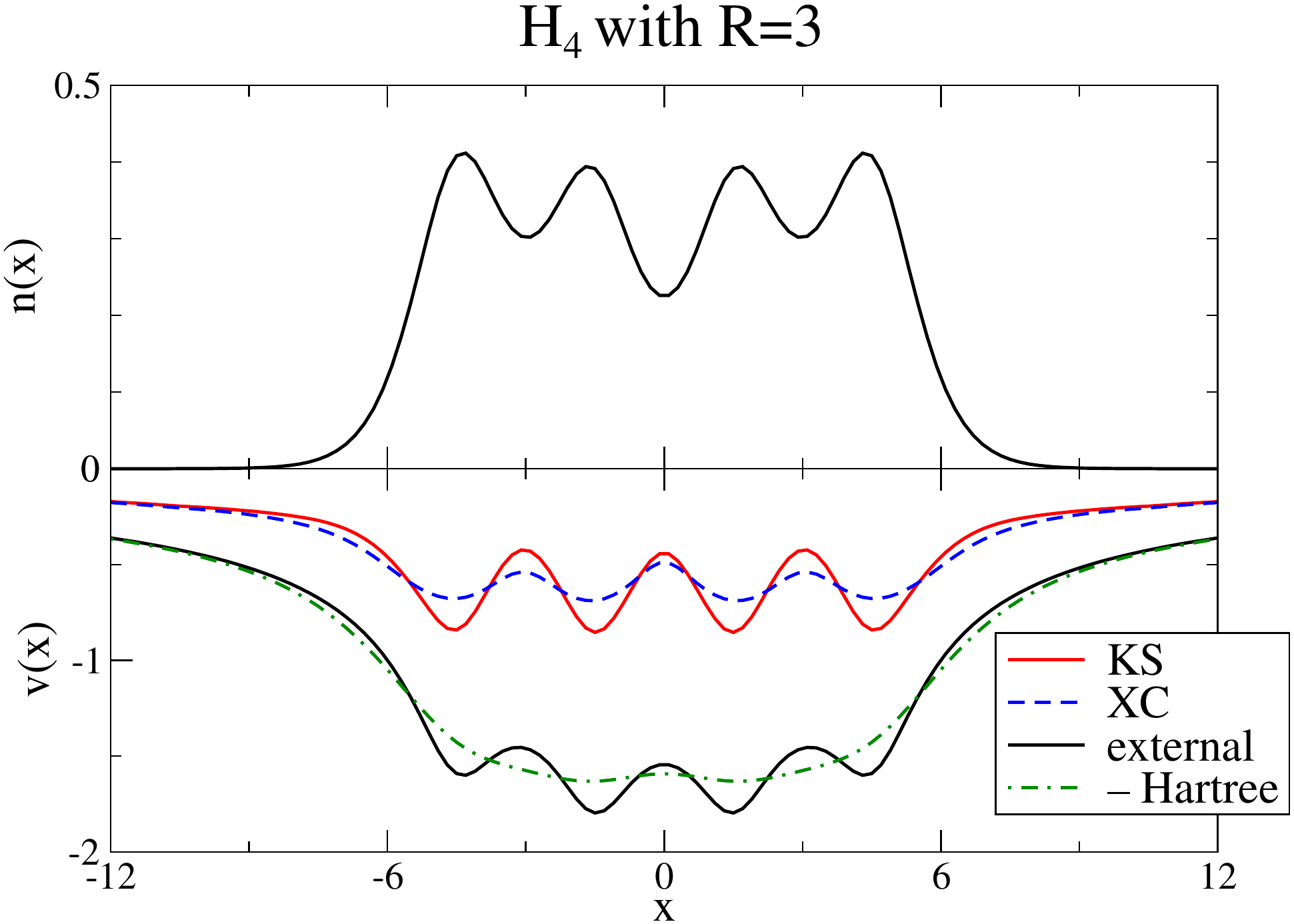}
\caption{External, KS, Hartree, and XC potentials, as well as the ground-state
density, for a moderately correlated 4-electron system
(four hydrogen atoms separated by an interatomic spacing of $R=3$). }
\label{H4b3xc}
\end{figure}

In \Figref{H4b3KS}, we plot the trial densities and KS potentials
for a four-electron, four-atom system.  The interatomic spacing
$R$ is chosen to make correlations moderate.
Choosing a density mixing of
$\lambda = 0.30$ affords fairly  rapid convergence.
We find that the final density, calculated within our KS algorithm, is equal
to the true ground-state density of the system.
We plot the final converged KS, Hartree, and XC potentials in \Figref{H4b3xc}.

Regarding stationary points of the exact functional,
we find that, in all the cases we ran, our KS algorithm 
converged to the true ground-state density.
An analytic result confirms that, given $v$-representable densities,
the only stationary point of the exact
KS scheme is the ground-state density of the system \cite{PL85}. 
We can see this by plugging the 
exact $v\Hxc[n](\br)$ from \Eqref{eqn:vHxc} into the KS update \eqref{eqn:KSupdate}.
The exact scheme then proceeds as
\ben
v\s(\br) := v\s[n_\text{in}](\br) + \big( v(\br) - v[n_\text{in}](\br) \big),\label{eqn:exactKSupdate}
\een
with self-consistency reached when $v(\br) = v[n_\text{in}](\br)$.  
This occurs at precisely one density:
at the ground-state density $n\gs(\br)$,
which is unique by the HK theorem.  
Thus the exact KS scheme has only
one stationary point for $v$-representable densities.

In density functional theory, there is no guarantee that a KS potential exists 
for a given physical system.  The guarantee is that if it does exist, it is unique and, as we pointed
out above, the only stationary point of the KS equations.  
Densities with legitimate KS potentials are called non-interacting $v$-representable.
We have performed many non-interacting inversions on accurate ground-state
densities of atomic chains, and have always found their KS potentials to exist, even
when the bond lengths are stretched.   
Since standard density functional approximations
usually become inaccurate for strongly correlated systems, such as when bonds are stretched,
a potential pitfall for KS-DFT is that such systems may fail to be non-interacting $v$-representable.
While there are subtleties to identifying whether a density is $v$-representable or not
(as discussed further in \Secref{s:nonvrep}), 
$v$-representability does not appear to be the main issue when strong correlation is involved \cite{Pc85,BBS89,HTR09,TMM09}. 
Instead, good approximate functionals simply are missing at present \cite{FNGB05,GB06}. 
If $v$-representability were to blame, the entire KS apparatus, despite being exact in principle, could
not be applied to such systems.  Happily, our results show no
evidence of such a disastrous situation.

\ssec{First steps}\label{s:conv}

Knowing that there is only one stationary
point of the KS scheme (for $v$-representable densities) tells us
nothing about the difficulty in finding it.
In this section we consider the most basic part of the KS scheme --
a single step in the KS algorithm -- which will help us understand
the convergence behavior of the exact functional for different systems.
We will see why strongly correlated systems
are more difficult to converge than weakly correlated systems.

\begin{figure}
\includegraphics[width=\columnwidth]{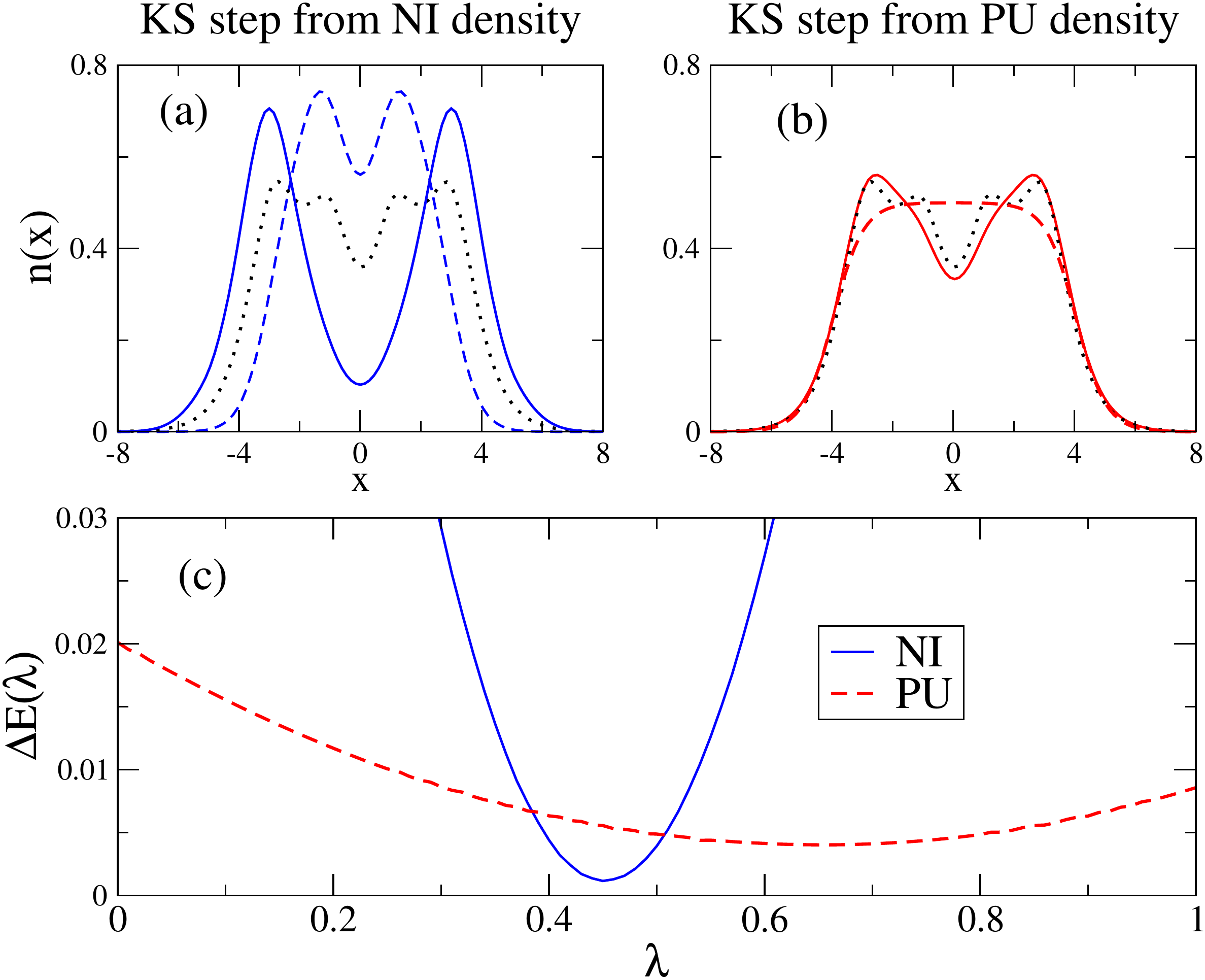}
\caption{
A single step in the KS scheme for a weakly correlated system 
(H$_4$ with $R=2$)
away from two different initial densities:
non-interacting electrons in the external potential (NI)
and a pseudouniform electron density (PU).
These initial densities are the dashed curves in (a) and (b),
and the solid curves are the output densities for each KS step;
for comparison the dotted curve is the exact density.
The lower panel plots \Eqref{eqn:DeltaE}, the energy of the 
system as it interpolates from the input to the output density.
 }
\label{H4b2-onestep}
\end{figure}

\begin{figure}
\includegraphics[width=\columnwidth]{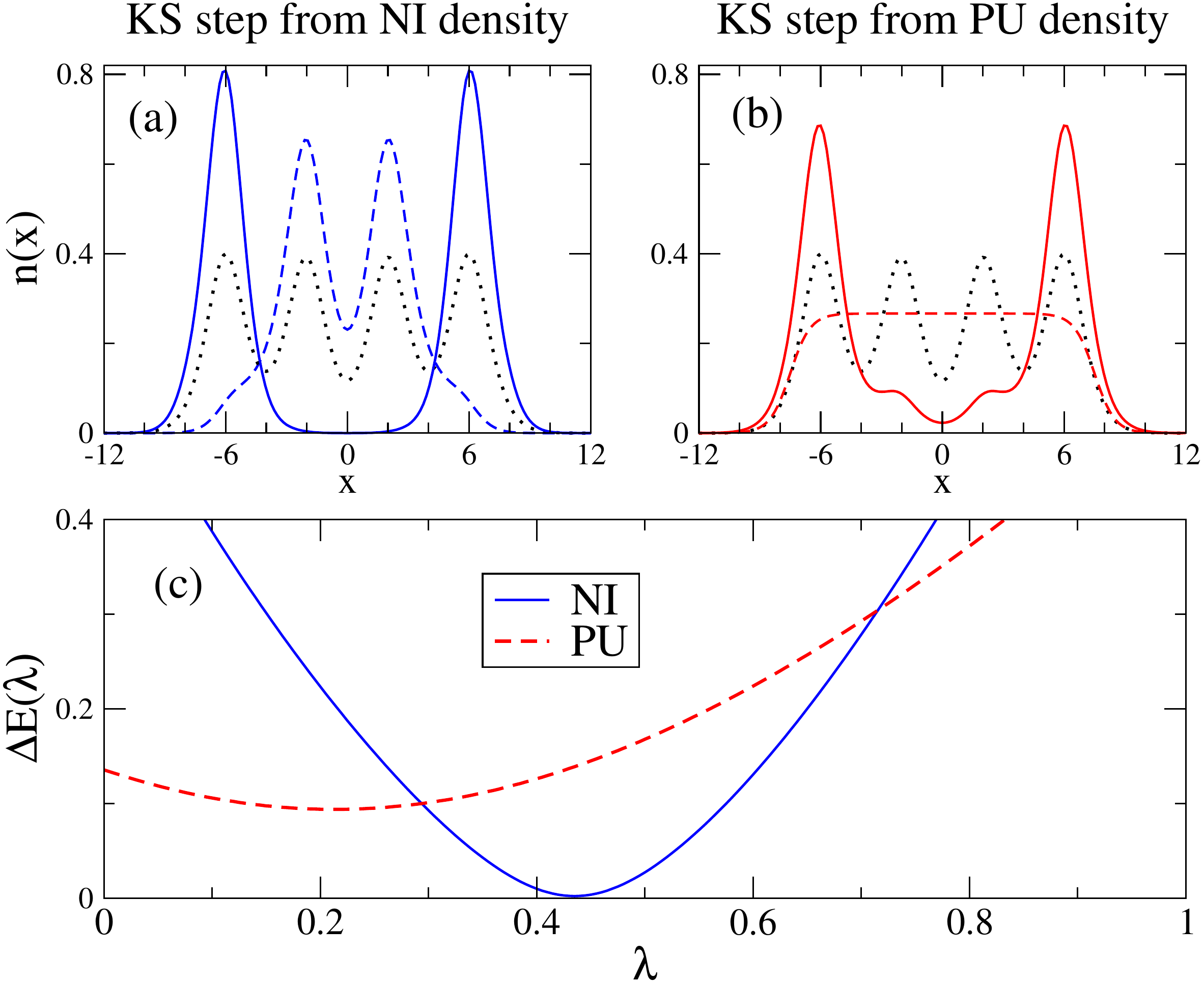}
\caption{
A single step in the KS scheme for a strongly correlated system 
(H$_4$ with $R=4$)
away from two different initial densities:
non-interacting electrons in the external potential (NI)
and a pseudouniform electron density (PU).
These initial densities are the dashed curves in (a) and (b),
and the solid curves are the output densities for each KS step;
for comparison the dotted curve is the exact density.
The lower panel plots \Eqref{eqn:DeltaE}, the energy of the 
system as it interpolates from the input to the output density.
 }
\label{H4b4-onestep}
\end{figure}

To explore how the KS scheme converges, we calculate the energy of the system 
which interpolates between the input and output densities for a single step of the algorithm,
measured against the ground-state energy:
\ben
\Delta E(\lambda) = E_v[n_\lambda] - E_v,\label{eqn:DeltaE}
\een
where $n_\lambda(\br)$ linearly interpolates between the input density (at $\lambda = 0$) to
the output density ($\lambda = 1$), just as in \Eqref{eqn:densmix}.
We plot $\Delta E(\lambda)$ as well as the input, output, and exact densities for various systems
in Figures~\ref{H4b2-onestep} and \ref{H4b4-onestep}.
As can be seen, the output density is in the right direction to minimize $E_v[n]$,
but it overshoots the minimum. 
Starting the next iteration of the KS scheme with
this output density would not (in general) 
allow convergence; therefore a mixture of
the  input and output densities is used as the next input,
thus motivating \Eqref{eqn:densmix}.
The optimal mixing $\lambda$ minimizes
$E_v[n_\lambda]$ on the interval $(0,1]$, and could be found using a line search.
But even with the optimal mixing, neither of the chosen starting points (a non-interacting and a
pseudouniform density) produces the ground-state density on the first iteration,
so it takes a few iterations to converge.
It is perhaps surprising, however, that a single iteration of the KS scheme
could get so close to the ground state.  For the weakly correlated system
(\Figref{H4b2-onestep}), the non-interacting starting point
gets within $\Delta E = 0.001$ of the ground-state energy with $\lambda = 0.45$, whereas 
the pseudouniform starting point minimizes $\Delta E = 0.004$ with $\lambda=0.45$.
For the strongly correlated system (\Figref{H4b4-onestep}), the optimal $\lambda$'s are smaller
and the $\Delta E$'s are larger:
the non-interacting initial point
minimizes at $\Delta E = 0.002$ with $\lambda = 0.44$, and the 
pseudouniform initial point minimizes $\Delta E$ at 0.094 around $\lambda = 0.21$.

Figures \ref{H4b2-onestep} and \ref{H4b4-onestep}
each plot only two cuts through the infinite-dimensional functional landscape. 

Figure \ref{H4b2-onestep} models a weakly correlated system---a four atom system
with an interatomic spacing of $R=2$---where a Slater determinant \cite{S29}
of non-interacting electrons is a good approximation to the underlying wavefunction.
But as we stretch the bonds to $R=4$ for \Figref{H4b4-onestep}, 
strong static correlation arises, and the KS wavefunction is
less like the true wavefunction of the interacting system
than that of \Figref{H4b2-onestep}.  Thus the density of a non-interacting system
in the external potential is a poor start for the KS scheme, and
energy differences from the ground-state are larger for 
the strongly correlated system than for the weakly correlated system.
Besides the scale, one might ask how the functional landscape 
differs between strongly correlated systems and weakly correlated systems.
While the two NI curves in Figures \ref{H4b2-onestep} and \ref{H4b4-onestep}
are deceptively similar, the PU curves begin to reveal
the treacherous landscape of the strongly correlated system near the 
minimum.

\begin{figure}
\includegraphics[width=\columnwidth]{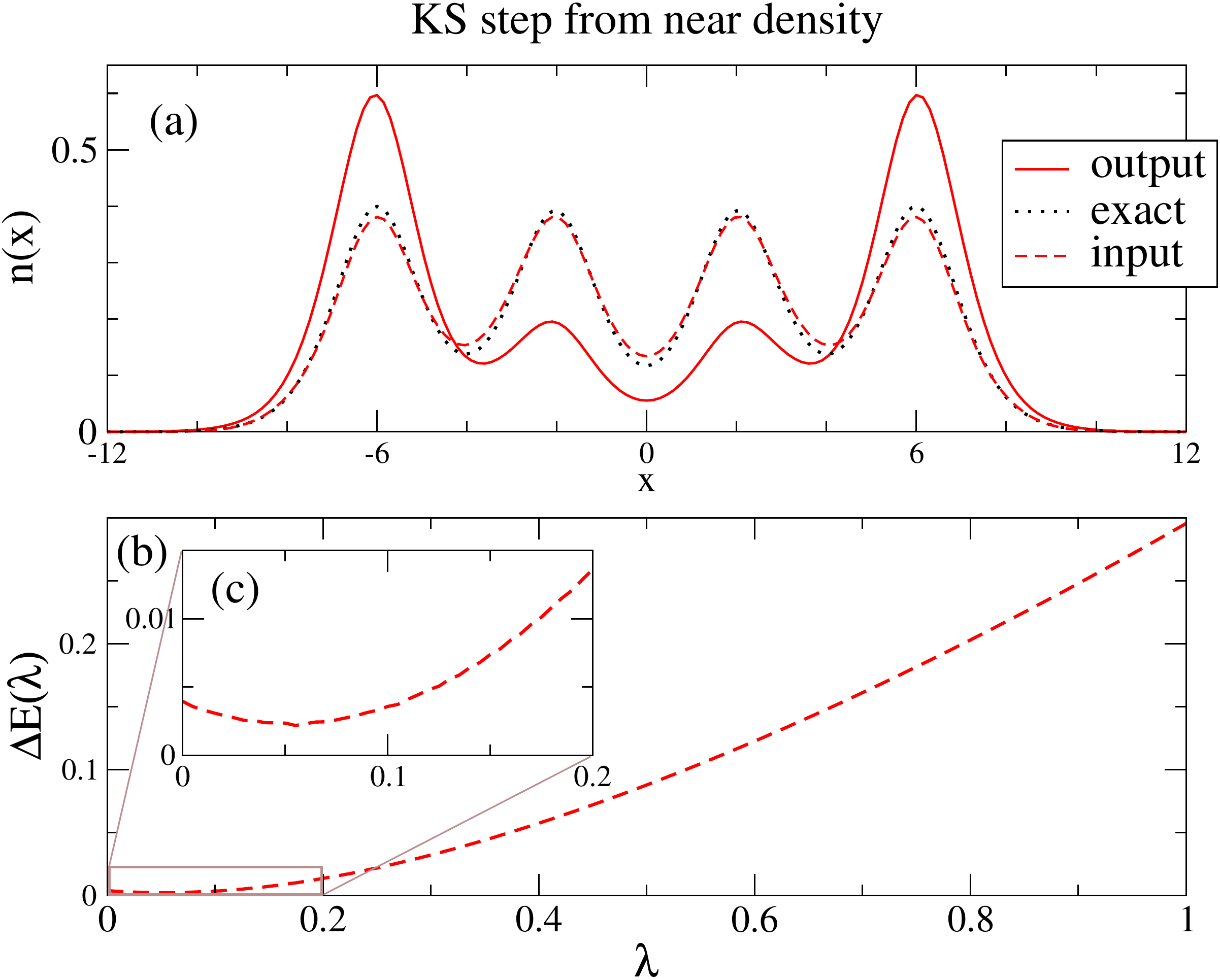}
\caption{
Taking a second step in the KS scheme for a strongly correlated system
(H$_4$ with $R=4$).  Panel (a) shows the input density which is
near to the exact density (the $\lambda = 42\%$ density of the 
NI input density of \Figref{H4b4-onestep})
and the resulting output density, which is far from the ground-state.
The lower panel (b) plots \Eqref{eqn:DeltaE},
and the inset (c) magnifies the small $\lambda$ region.
 }
\label{H4b4-onestep2}
\end{figure}

We now look at the second iteration of the KS scheme to see if there
is a difference between the strongly and weakly correlated systems.
We choose the NI-path
density from \Figref{H4b4-onestep} with a good (but not optimal) 
mixing of $\lambda=42\%$ as input into the KS equations.  
For the weakly correlated system of \Figref{H4b2-onestep}, the
second KS step (not shown) looks much like the first step, though
with a much smaller energy scale involved.  Thus a fairly large $\lambda$
may be used when correlations are weak, and convergence is rapid.
But it is not the same for the strongly correlated system.
 As shown in \Figref{H4b4-onestep2}, the next
iteration of the KS procedure will not allow us to make the same giant stride
as in the first iteration.  For the new $\lambda$-mixed density, 
we again evaluate $\Delta E(\lambda)$ from \Eqref{eqn:DeltaE} and find
that it reaches a minimum much sooner.  
Thus a much smaller $\lambda$---around 6\% as seen in the inset---must 
be chosen in order not to go far off track.
Furthermore, choosing even the optimal $\lambda$ does not result in a much better
energy  as it did in the first iteration.
This  makes convergence a long and difficult process, since we can only afford
to take small steps.

In the last part of this section, we give some formulas which may aid in
determining the optimal $\lambda$ each step.
We consider derivatives of $E_v(\lambda) \equiv E_v[n_\lambda]$ with respect to $\lambda$.
For example, large $E_v''(\lambda) \equiv d^2 E_v[n_\lambda] / d \lambda^2$ 
relative to the magnitude of $E_v'(\lambda) \equiv d E_v[n_\lambda] / d \lambda$ requires
a smaller $\lambda$ to lower the energy.  
Given some bound on $E_v''(\lambda)$, one could analytically determine a safe (i.e.\
not too large or too small) approximation to the optimal $\lambda$ \cite{A66}.   
The derivatives of $E_v(\lambda)$ may be taken analytically \cite{U12,WSBW13}: 
\begin{widetext}
\begin{eqnarray}\label{eqn:1stfctderiv}
E_v'(\lambda)&=&\left.\intr \frac{\delta E_v[n]}{\delta n(\br)}\right|_{n_\lambda(\br)}
			\big(n_1(\br)-n_0(\br)\big) 
=\intr \Big(v(\br)+v\Hxc[n_\lambda](\br)-v\s[n_\lambda](\br)\Big)
		 \big(n_1(\br)-n_0(\br)\big)\\
E_v''(\lambda)&=&\intrrp \big(n_1(\br)-n_0(\br)\big)
	\Big(f\Hxc[n_\lambda](\br,\br')-\chi\s^{-1}[n_\lambda](\br,\br')\Big)
		\big(n_1(\br')-n_0(\br')\big),
\label{eqn:2ndfctderiv}
\end{eqnarray}
\end{widetext}
where $n_1(\br) = n_\text{out}(\br)$ and $n_0(\br) = n_\text{in}(\br)$ for the 
current KS step of interest, the HXC kernel $f\Hxc[n](\br,\br')$ is:
\ben
f\Hxc[n](\br,\br') = \chi\s^{-1}[n](\br,\br') - \chi^{-1}[n](\br,\br'),
\een
 and $\chi\s^{-1}[n](\br,\br') = \delta v\s[n](\br) / \delta n(\br')$ 
($\chi^{-1}[n](\br,\br') = \delta v[n](\br) / \delta n(\br')$) is the non-interacting (interacting)
inverse density-density response matrix.  
Calculating $f\Hxc[\n](\br,\br')$ is quite challenging, and has recently been evaluated
with time dependence for some simple systems \cite{TK14}.

We emphasize that $n_\text{out}(\br)$ is a functional
of $n_\text{in}(\br)$ and does not depend on $\lambda$ at all.
Thus \Eqsref{eqn:1stfctderiv} and \eqref{eqn:2ndfctderiv} are strictly functionals of the
input density $n_0(\br)$ alone.

Towards the end of approximating the optimal $\lambda$, one may fit 
$E_v[n_\lambda]$ given some information on the derivatives.
At the end points the derivatives simplify to
\begin{eqnarray}
E_v'(0)&\equiv&\intr \Big(v\sone(\br)-v\szero(\br)\Big)\big(n_1(\br)-n_0(\br)\big)\\
E_v'(1)&\equiv&\intr \Big(v\Hxcone(\br)-v\Hxczero(\br)\Big) \big(n_1(\br)-n_0(\br)\big),\quad \ \
\end{eqnarray}
where $v\sj(\br) = v\s[n_j](\br)$ and $v\Hxcj(\br) = v\Hxc[n_j](\br)$.
We find that in many systems a Hermite spline fit \cite{NumRec} (using $E_v(0)$, 
$E_v(1)$, and the
derivatives $E_v'(0)$ and $E_v'(1)$) is a good approximation to the
energy curve $E_v(\lambda)$, or at least to where it attains the minimum.  
However, this fit requires an inversion to find
$E_v(0)$ and $E_v'(0)$, which may be impractical for standard KS calculations.

\ssec{Why convergence is difficult for strongly correlated systems}\label{s:diff}

In this section, we discuss an important reason why
convergence is difficult for strongly correlated systems, and mention  some algorithms which
counteract the underlying problem.
Frequently, systems with strong static correlation
possess a small gap \cite{GS10}, which in turn makes convergence difficult \cite{DS00}.
We can understand this difficulty by considering the non-interacting density-density
response matrix $\chi\s(\br,\br')$:
\ben
\chi\s(\br,\br') = 2 \sum_{i\neq j}^\infty \dfrac{ f_j - f_i }{ \epsilon_j - \epsilon_i }
\phi_i(\br) \phi^*_j(\br) \phi_i^*(\br') \phi_j(\br'),
\een
where $0 \le f_j \le 1$ is the Fermi occupation of orbital $\phi_j(\br)$.  
For a small gap system, $\epsilon\LUMO - \epsilon\HOMO$ is particularly small,
making that term in $\chi\s(\br,\br')$ especially large.
This means that small changes in the KS potential can produce
large changes in the density, which makes convergence in the KS scheme difficult.
We can visualize this property by performing a singular value decomposition on $\chi\s(\br,\br')$,
as in \Eqref{chiSVD}.  Equivalently, since $\chi\s(\br,\br')$ is symmetric
in $\br, \br'$, we can diagonalize $-\chi\s(\br,\br')$:
\ben
\chi\s(\br,\br') = -\sum_{\beta=1}^\infty a_\beta M_\beta(\br) M_\beta(\br'), \label{chisSVD}
\een
where $M_\beta(\br)$ ($a_\beta$) are the eigenvectors (eigenvalues) of $-\chi\s(\br,\br')$.
Since $\chi\s(\br,\br')$ is negative definite, we can order $a_{\beta} \ge a_{\beta+1} > 0$.
The breakdown in \Eqref{chisSVD} physically means that
a change in the KS potential along the direction $-M_\beta(\br)$ 
produces a change in the density along $M_\beta(\br)$ with a magnitude given by $a_\beta$,
at least to first order.
We therefore call $M_\beta(\br)$ the density response vectors and $a_\beta$ the response
amplitudes of $\chi\s(\br,\br')$.
The amplitudes depend on the normalization of $M_\beta(\br)$, and the standard squared ($L^2$) norm is
not the most natural choice.
Because $M_\beta(\br)$ corresponds to a change in density, 
we choose $\intr |M_\beta(\br)| = 2$ so that $M_\beta(\br)$ can be thought of as moving
an electron from one region (where $M_\beta(\br) < 0$) to another (where $M_\beta(\br) > 0$).
Finally, because $a_\beta$ are ordered by importance, $\chi\s(\br,\br')$ can be accurately and 
efficiently represented by truncating
the sum once $a_\beta$ drops below some tolerance.

\begin{figure}
\includegraphics[width=\columnwidth]{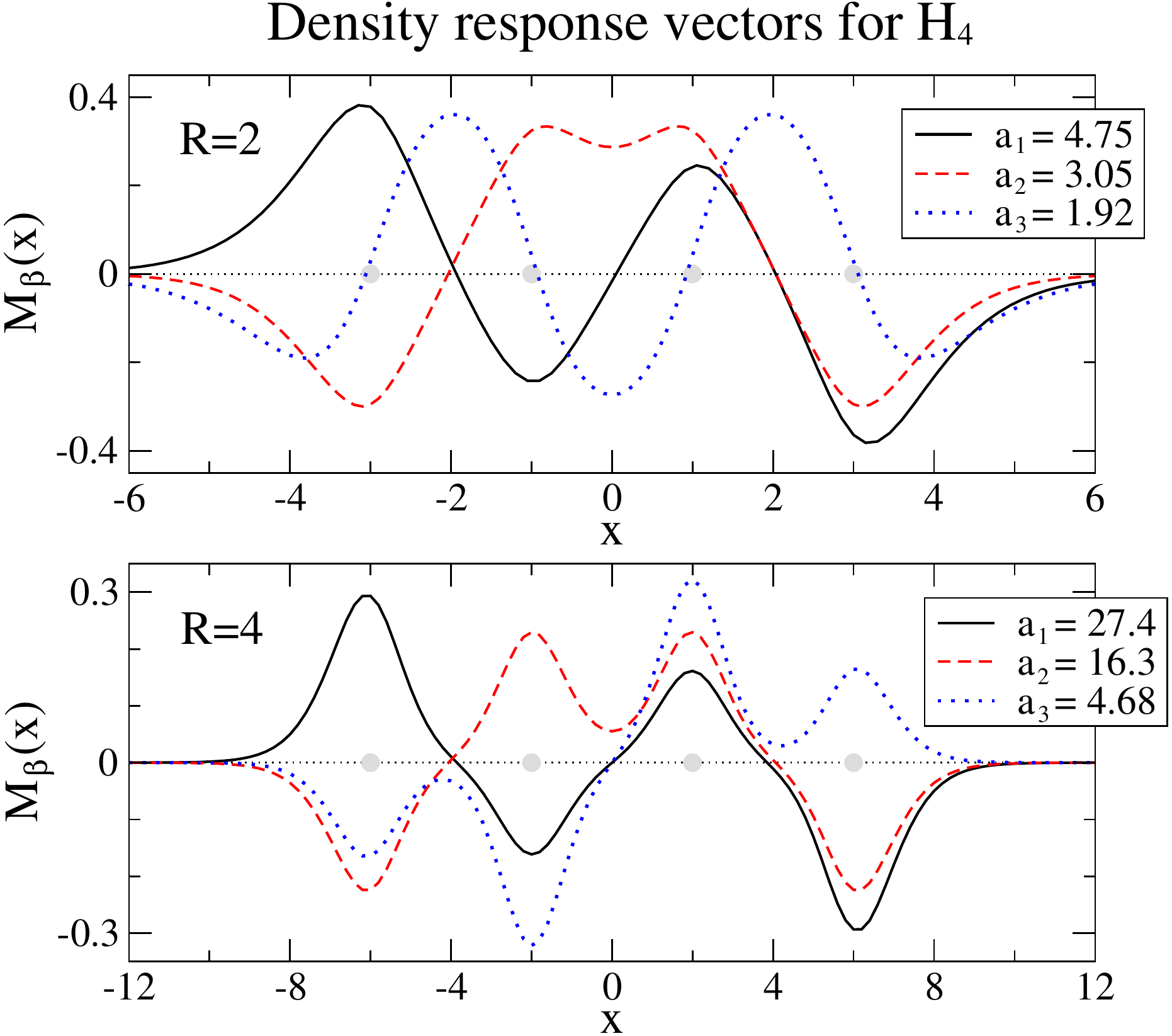}
\caption{
The most important density response functions $M_\beta(x)$ from \Eqref{chisSVD}
and their response amplitudes $a_\beta$, for the weakly correlated system ($R=2$) in the top panel,
and the strongly correlated system ($R=4$) in the bottom panel.  The locations
of the atoms are shown in solid grey circles.
}
\label{H4chis}
\end{figure}

We can easily find the density response vectors
$M_\beta(x)$ for the 1d H$_4$ systems we have already discussed at length,
which allows us to diagnose our convergence difficulties.
In \Figref{H4chis}, we plot the first few most important $M_\beta(x)$.  The first
two ($\beta=1,2$) look similar for the weakly correlated and the strongly correlated
systems, though the response amplitudes $a_\beta$ are quite different.  If the potential
changes in the direction $-M_1(x)$, it drives a strong density response
in the direction $M_1(x)$ due to the large response amplitude $a_1 = 4.75$ at $R=2$ and $a_2=27.4$ at $R=4$.
Luckily, we can assume reflection symmetry, so that
in the iteration of the KS equations we do not have to worry about contributions
from these $\beta = 1$ terms.  But now consider the symmetric $\beta =2$ terms.  
If the KS potential changes in the direction $-M_2(x)$, 
the density will respond by changing in the direction $M_2(x)$, and the response amplitude
is very strong for the $R=4$ system ($a_2 = 16.3$).

These (ground-state) response properties  can be used to
explain the problems that we have converging the strongly correlated H$_4$.
If the initial KS
potential puts most of the density around the central two atoms, to 
compensate the next trial
KS potential \eqref{eqn:KSupdate} will increase in the central region and decrease for the edge atoms.
In response,
the new density will place too many electrons on the edge atoms.
We have already seen this in Figures~\ref{H4b2-onestep} and
\ref{H4b4-onestep} with the NI starting densities.  
The reverse can also happen, where
most of the input density is on the edge atoms, and
 the output density is more centralized.  
For the strongly correlated H$_4$,
this ``sloshing" back and forth can be particularly strong because the response amplitude
$a_2$ is quite large  -- 
this problem plagues densities even very close to the ground state,
as seen in \Figref{H4b4-onestep2}. As $R\to \infty$, $a_2$ diverges,
making it more and more difficult to converge.
To ameliorate these problems, some convergence schemes artificially increase
the gap \cite{SH73} or populate otherwise unoccupied orbitals \cite{RS99b}. 
For other discussions on this matter, see \Ref{CM13} and for implications
for time-dependent DFT, \Ref{Mc05}.

\ssec{Convergence as correlations grow stronger}

In this section, we explore convergence within the simplest density
functional approximation, the local density approximation (LDA) \cite{KS65}, 
in order to understand some basic limits on convergence as well as
its dependence on the KS gap, i.e.\ the HOMO-LUMO gap. A simple expression for the LDA is available for our model 1d systems \cite{HFCV11,WSBW12}.
We expect the LDA to converge in a similar way to
the exact functional, especially when the KS gap of the system is close 
for both self-consistent LDA and exact solutions \cite{NSb84}. 
We therefore use it to study more broadly the convergence behavior of the KS scheme
applied to H$_2$ with variable bond length.
As before, changing the bond length allows us to tune the strength of the correlation:
at small bond lengths the system is weakly correlated and at large bond 
lengths strong static correlation arises \cite{WSBW12}. 
To aggravate convergence difficulties, we choose the
 initial density to be entirely centered on
 one atom \cite{WSBW13}, and determine the $\lambda$ values for
which the KS scheme will converge, as well as how quickly.
Furthermore, we enforce spin-symmetry, so while the restricted LDA
energy is wrong in the $R\to\infty$ limit \cite{WSBW12},  
we expect to see convergence behavior similar to the exact
functional \cite{WSBW13}.

\begin{figure}
\vspace{0.5cm}
\includegraphics[width=\columnwidth]{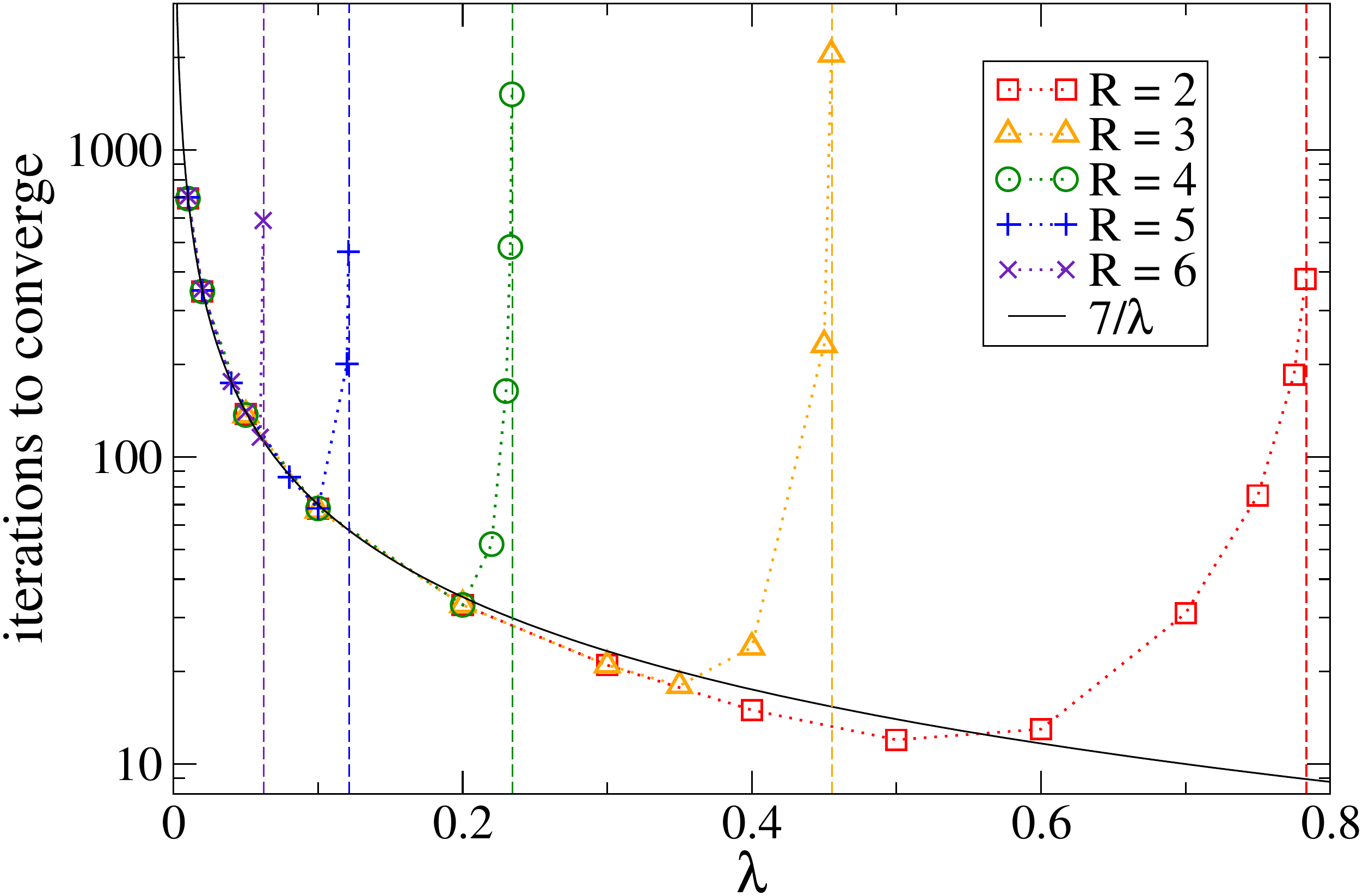}
\caption{The number of iterations required to converge an LDA calculation
to $\eta < 10^{-8}$ \eqref{eqn:tolerance}, as a function of $\lambda$, for
various bond-lengths $R$ of the H$_2$ molecule, starting with an initial
density of H$^-$ on the left atom.  The asymptotic form for small
$\lambda$ can be well-approximated by $7/\lambda$ for the data shown.
 }
\label{LDAiterationsH2}
\end{figure}

In  \Figref{LDAiterationsH2}, we plot the number of iterations required to
converge an LDA calculation to $\eta < 10^{-8}$ as a function of $\lambda$,
for a variety of bond lengths $R$.  
Each curve ends at $\lambda_c(R)$, the largest $\lambda$ for which 
the damped KS algorithm converges from this initial density.
For a weakly correlated system (e.g.\ $R=2$),
a very large $\lambda$ will produce convergence, and the optimal $\lambda$
to converge in the fewest iterations is also fairly large (around $0.5$ for $R=2$).  
As the bond length is stretched, both the critical $\lambda$, $\lambda_c(R)$, as
well as the optimal $\lambda_O(R)$ decrease.  
In response, the minimum number of iterations  $N_\text{min}(R)$ to converge to a tolerance
$\eta< 10^{-8}$,
increases.  For example, $N_\text{min}(R=2) = 12$ for $\lambda_O(R=2) \approx 0.5$.
Considering the iterations it takes to converge as a function of $\lambda$,
we see that as $\lambda$ decreases past the optimal $\lambda$,
it begins to take longer to converge the calculation.
For $\lambda \rightarrow 0$, we approach an asymptote that appears valid
for all values of $R$, given this initial starting point in the H$_2$ system:
$N_\text{asym}(\lambda) = 7/\lambda$.  While this is by no means a universal asymptote for all systems,
we recognize there is a fundamental limit to how quickly we can converge as $\lambda \to 0$.

\begin{figure}
\includegraphics[width=\columnwidth]{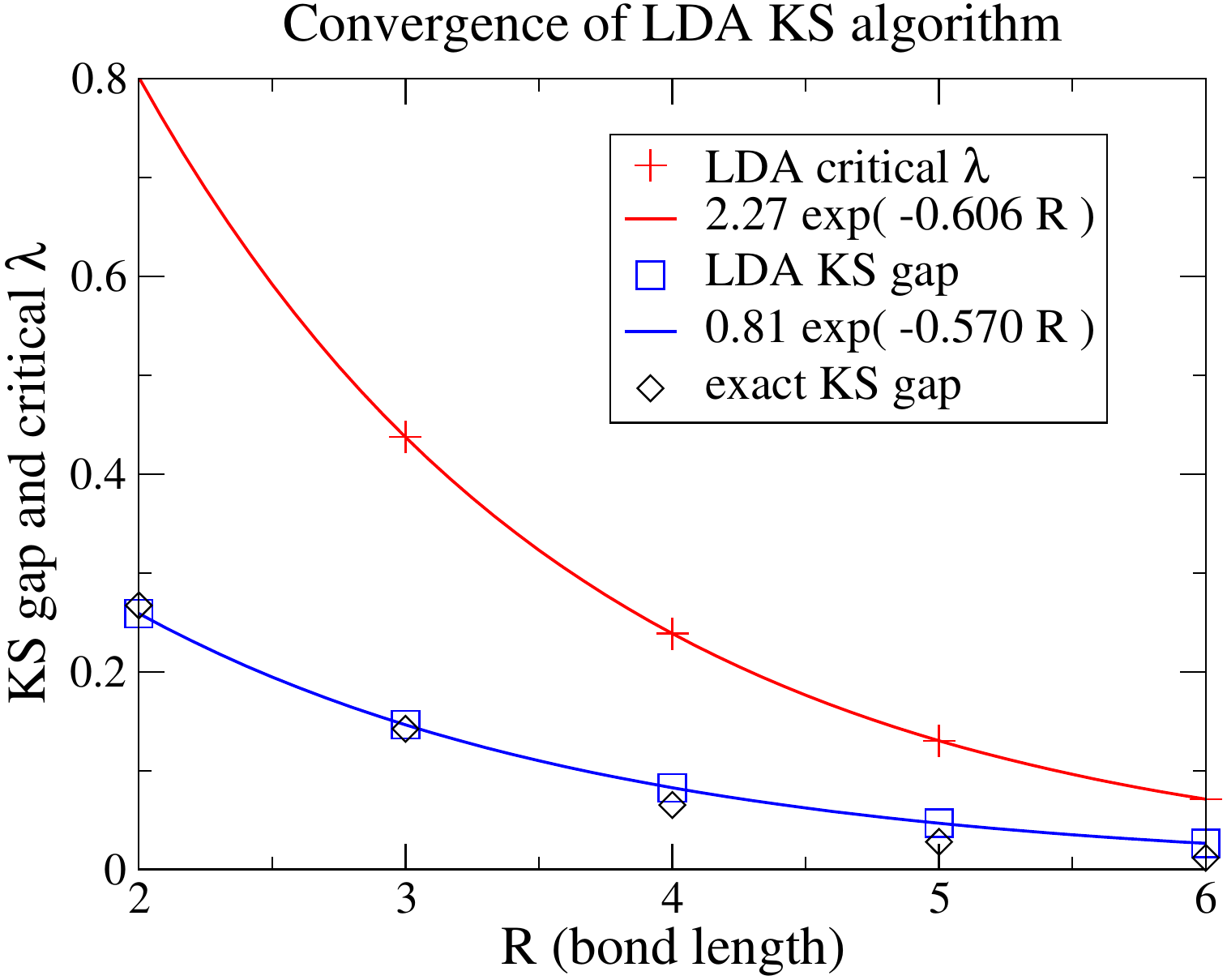}
\caption{
Plotting $\lambda_c$ for an LDA calculation,
as a function of the bond length $R$ of a stretched
hydrogen molecule, starting with the exact H$^-$ density on one atom 
as well as KS gaps for both the LDA and exact systems.
 }
\label{blambda}
\end{figure}

In \Figref{blambda}, we plot the convergence-critical $\lambda$ value 
 as a function of the bond length $R$, as well as the KS gap of both the LDA
and exact systems.  The LDA KS gap decays at about the same
rate as the critical $\lambda$, an observation that makes sense given that
the KS gap has such an important role in convergence -- the smaller
the gap the more difficult it is to converge the calculation \cite{NSb84}. 
For bond lengths $R \lesssim 4$, the LDA KS gap is quite close
to the exact KS gap, so that we expect similar convergence behavior for the exact
functional.  However, as $R$ increases, the true KS gap decays more quickly than the LDA KS gap,
so that the exact calculation has an even greater difficulty converging \cite{WSBW13}.   
It could be that some values of $\lambda$ larger than $\lambda_c$
allow for convergence if the density fortuitously lands close enough to the ground state in some iteration,
but there is no systematic approach to find these $\lambda$.

\ssec{Classifying convergeability}

In this section, we want to mathematically investigate the
the space of densities that allow convergence and how quickly that occurs.  That is,
given some initial density and a fixed value of $\lambda$, can we determine whether the KS
scheme will converge within some given number of iterations?
With $\lambda$ too large, the KS scheme will be doomed to repeatedly overstep the
ground-state density.  

To quantify these ideas,
define $\eta^M[n](\lambda)$ to be the value of $\eta$ defined by Eq.~\eqref{eqn:tolerance}
after $M$ iterations of the KS equations with a fixed
mixing of $\lambda$, starting with the input density $n(\br)$.  Then 
define the density set:
\ben
S_\zeta^M(\lambda) 
\equiv \big\{ n(\br) \text{ s.t. } \eta^{M}[n](\lambda) < \zeta   \big\}.
\label{eqn:sets}
\een
This set describes the densities $n(\br)$ which converge to $\eta < \zeta$
in a finite number of iterations $(M)$, given a fixed-$\lambda$
iteration of the KS equations.  
For example,
$S^1_\zeta \equiv S^{1}_\zeta(\lambda=1)$ is the set of input densities $n_\text{in}(\br)$ 
that are within $\eta < \zeta$ of their output densities.  (For
one step, $\lambda$ does not matter.)
This set \eqref{eqn:sets}
allows us to quantify the different levels of convergence hell.
$S_\zeta^1$ is the lowest level, and includes the ground-state density.
$S_\zeta^2(1)$ is the second level, and also includes the ground-state
density. 
As $M$ becomes large (but remains finite), 
$S^{M}_\zeta(1)$ reaches out to the $M$th level of hell:
the set of densities which converge to within $\eta < \zeta$ within
a finite number of full-KS-step iterations.
All other densities belong to the
$\lambda = 1$ limbo density set, densities which are doomed to wander
for (essentially) all eternity, never to converge.
Similarly, there are less-strict convergence sets for $\lambda < 1$,
which describe a sort of density purgatory.

\begin{figure}
\begin{center}
\includegraphics[width=\columnwidth]{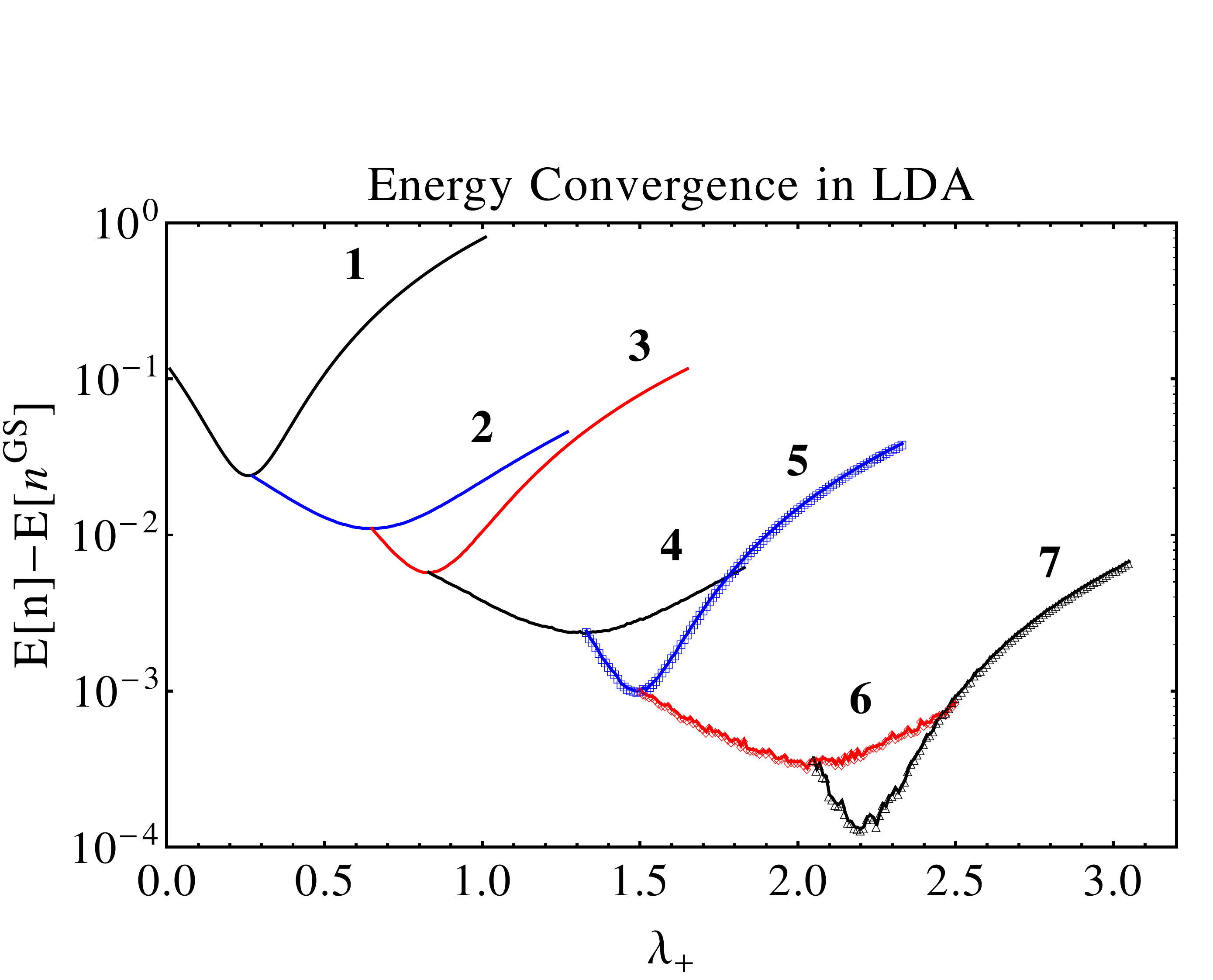}
\includegraphics[width=\columnwidth]{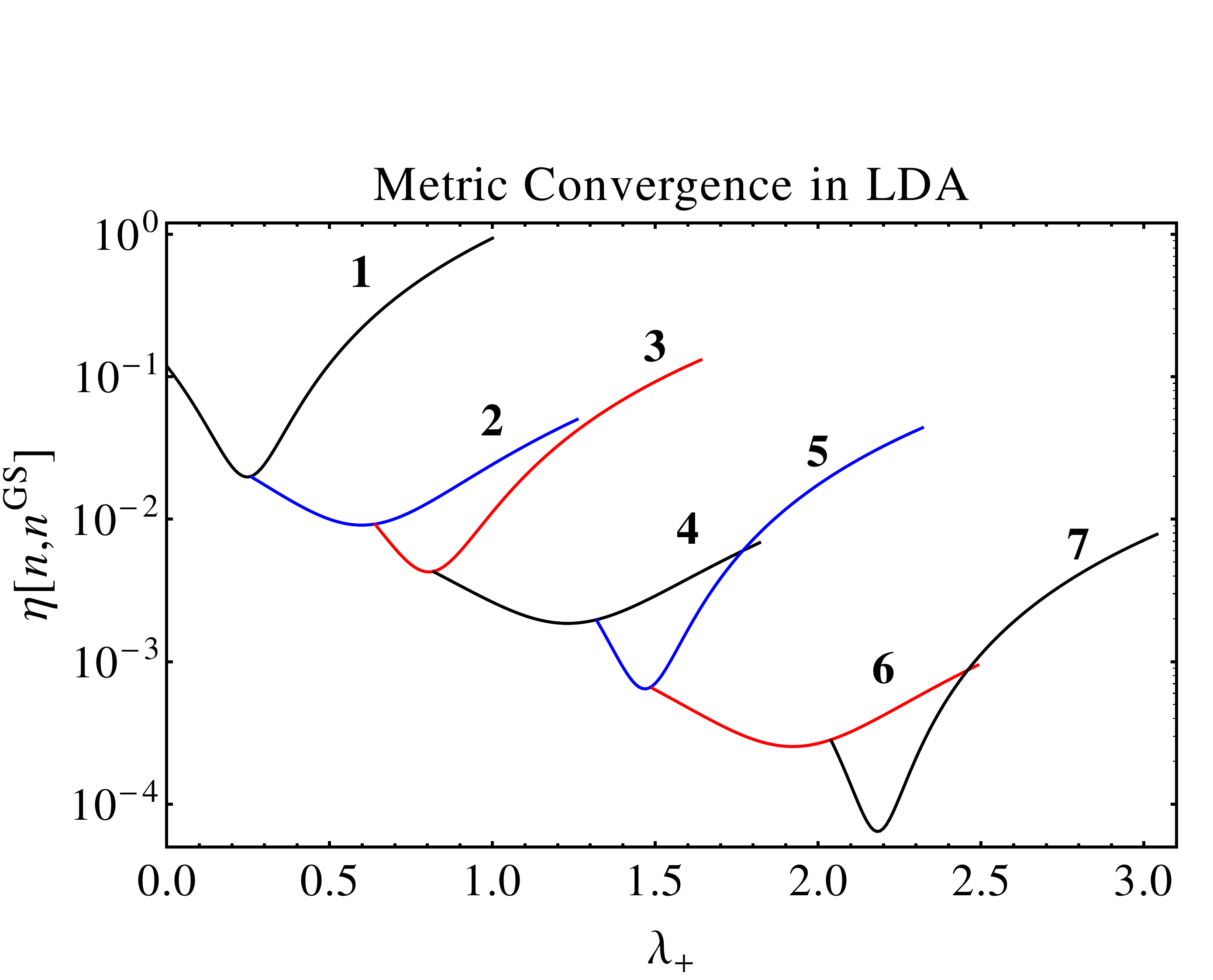}\\
\end{center}
\caption{
The first few steps (numbered) in the KS scheme from some arbitrary 
starting density for  H$_4$ with $R=4$ in the LDA approximation.  Numerical precision makes the energy data noisy. 
Metric distances are compared with the LDA ground state.  
Figure (a) plots the energy and (b) the metric as a function of  
the accumulated $\lambda$ step. 
The density $n_\lambda(x)$ with the lowest metric distance is not  the energetic minimum,
but they are fairly close.
}
\label{fig:metric}
\end{figure}

It might be hoped to connect these abstract convergence sets
with some concrete measure, say some metric between the ground-state density and 
the density inputted into the KS scheme, $\eta[n,n\gs]$.  Here we simply define the
metric similarly to our $\eta$ convergence quantifier:
\ben
\eta[n_1,n_2] = \intr \big( n_1(\br) - n_2(\br) \big)^2 / N^2.\label{etametric}
\een
The idea of a metric on the set of 
densities is not new \cite{DCFC11,M13}. 
Unfortunately, current metrics are not guaranteed to
correlate e.g.\ a given input density $n(\br)$ with a given convergence
set $S^{M}_\zeta(\lambda)$.  That is, there is likely no function $g^{M}_\zeta(\lambda)$
for which $\eta[n,n\gs] < g^{M}_\zeta(\lambda) \implies n(\br) \in S^{M}_\zeta(\lambda)$.
In \Figref{fig:metric} we show why.  For $\lambda_+$ the accumulated $\lambda$
throughout the KS scheme, we see that the metric $\eta[n_{\lambda_+},n\gs]$ tracks well 
with the how close the energy $E_v[n_{\lambda_+}]$ is to the ground state energy (at least
for this example, 1d H$_4$ in LDA).  Despite this
nice relationship between the energy and the metric, a small $\eta[n,n\gs]$ does not necessarily
mean we can take a large step in $\lambda$ each iteration.  Therefore we do not know
how many steps it will take nor how small a $\lambda$ is required based on the metric alone.
More physically motivated metrics might remedy this issue, but we must leave this
question open.

\ssec{Spin DFT}\label{s:spin}

In this section we extend the exact functional to include spin dependence.
We test the exact spin-dependent functional on the case of stretched H$_2$, starting our
KS scheme with a broken-spin-symmetry solution, to determine whether or not
the exact functional will find the correct spin-singlet ground state \cite{WSBW13}.

Treating the up-spin and down-spin electrons separately leads to much improved
density functional approximations, as well as new challenges \cite{BH72,VWN80}.  
If an unbalanced spin-state is provided as input to the KS scheme,
approximate spin-density functionals may find a broken spin symmetry
when the ground state should be a singlet.  This is the case for many
open-shell systems as bonds are stretched.
The simplest such system, and a paradigm of DFT failures,
is stretched H$_2$ \cite{CF49,PSB95,BAb96,WSBW12}. 
In this case, it is clear that the exact XC spin-density functional does not
break symmetry at the solution density, since the ground state of any two-electron
system is a singlet (in the absence of external magnetic fields) \cite{PSB95}.  This is true
of both the interacting wavefunction and the KS Slater determinant, which is then
just a doubly occupied molecular orbital.

To investigate these issues,
 we must first add spin-dependence to our functional, which is simple
enough in principle.
The added challenge is needing the ability to solve an interacting system with different
potentials for spin-up and spin-down electrons, i.e.\ electrons in a 
collinear magnetic field.  Similar to \eqref{eqn:vHxc},
the HXC potential for spin-$\sigma$ electrons is:
\ben
v_{{\sss HXC},\sigma}[n\up,n\dn](\br) = v\s[2n_\sigma](\br) - v_\sigma[n\up,n\dn](\br),
\een
where the KS potential for the up electrons can be inverted
independently of the down electrons by doubly
occupying the up density \cite{GBc04} (and vice versa for down electrons), 
and $v_\sigma[n\up,n\dn](\br)$ is the spin-$\sigma$ potential necessary to
produce spin densities $n\up(\br)$ and $n\dn(\br)$ from an interacting Hamiltonian.
We now investigate the use of the exact spin-dependent functional in
a system where standard approximate functionals have multiple stationary points.

\begin{figure}
\includegraphics[width=\columnwidth]{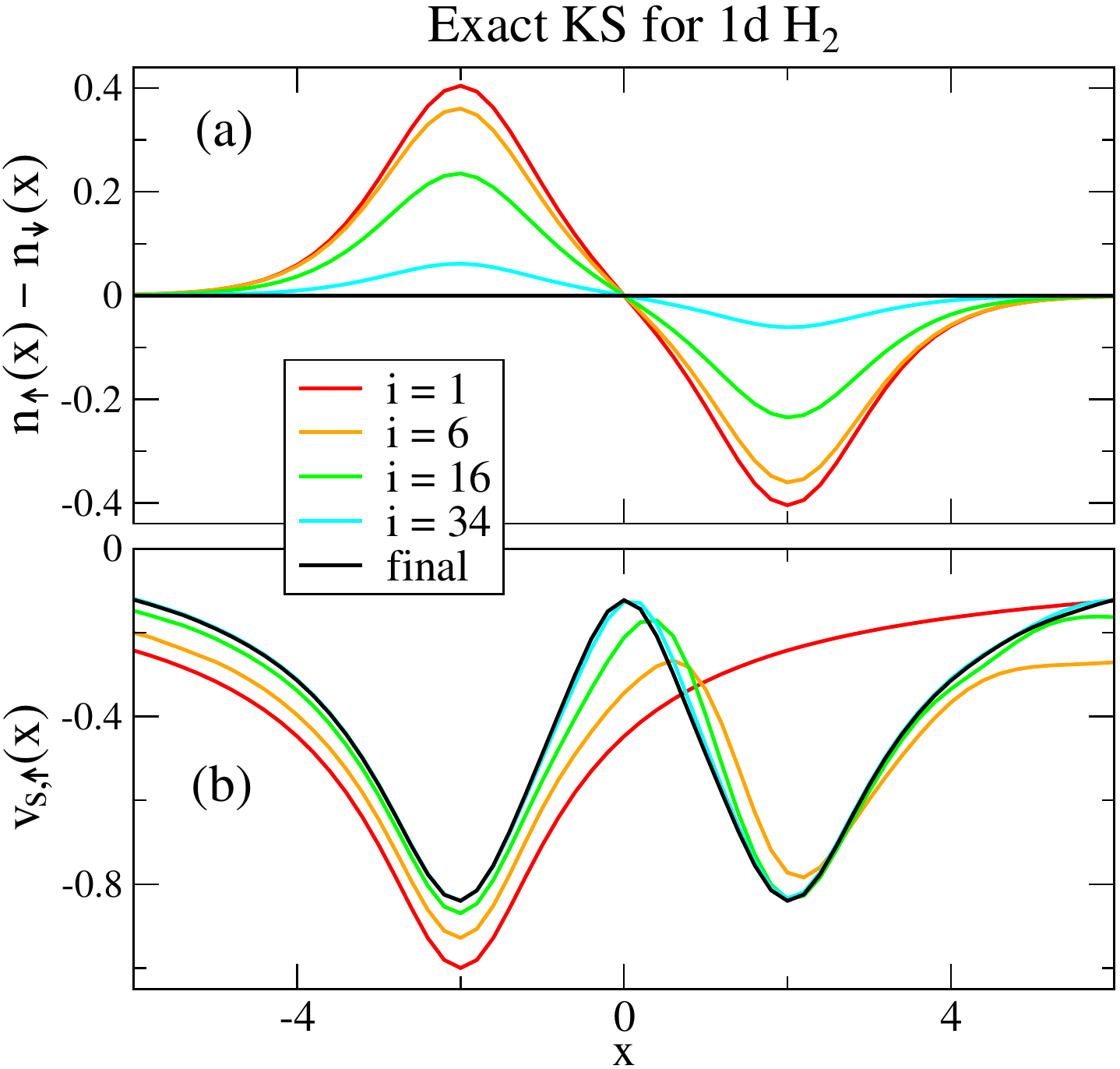}
\caption{
Starting an exact KS calculation of stretched H$_2$ with a spin-polarized 
density still converges to the correct spin-singlet density.
Through the iterations $i$, we plot (a) the polarization density $n_\uparrow(x) - n_\downarrow(x)$
and (b) the up KS potentials $v_{{\sss S},\uparrow}(x)$; the down potentials
are the mirror images.  
 }
\label{H2b4}
\end{figure}

To test whether the exact functional can find the singlet solution
for the stretched H$_2$ case, we start the exact KS
calculation with a spin-polarized initial density, with the
up electron on the left atom and the down electron on the right.
With this input, the KS scheme using the local spin-density approximation converges to
a broken symmetry solution \cite{WSBW12}.
But as seen in \Figref{H2b4},  the exact functional
finds the correct spin-singlet density without
much trouble.  
(For this system, a large density mixing was used, namely $\lambda = 50\%$.) 
So long as the spin-densities are $v$-representable, the arguments
of \Ref{PL85} apply, and there is only one stationary point of the
exact functional -- the true ground-state density.
This is true not only in 1d (as we have illustrated) but
also in 3d.

\ssec{Non $v$-representable densities}\label{s:nonvrep}

An important question that has haunted density functional theory since
the proofs of Hohenberg and Kohn is that of $v$-representability \cite{DG90}, i.e.,
for a given density $n(\br)$, does there exist a one-body potential $v[n](\br)$
for which it is the ground-state density?   The constrained-search formulation
of Levy \cite{L79} and of Lieb \cite{L83} bypasses this issue by defining  the functional $F[n]$
as an infimum over a given class of
wavefunctions.    
But our methodology of performing both interacting and non-interacting
inversions essentially requires $v$-representability in both the interacting
and non-interacting systems.  (In fact, $v\Hxc[n](\br)$ is ill-defined if 
$n(\br)$ is not $v$-representable \cite{L83,Leeuwen,L07}.)
In all our calculations to date, we have had no difficulty with $v$-representability,
but in the present section, we use explore its meaning in more detail.  

To be clear, we consider a density $v$-representable if
it is {\em ensemble} $v$-representable.
The generalization to mixed states (ensembles) is important
for degenerate systems, where not every density comes from a pure-state wavefunction \cite{V80,L82,L83,SGB97}; 
these practical details impact the calculations for and the values of the
functionals $F[n]$ and $T\s[n]$ \cite{L83,UK01,WSBW13}, but they are not our primary concern.
In addition, we focus on {\em non-interacting} $v$-representability;
the challenges for interacting $v$-representability are similar,
though the sets of interacting and non-interacting $v$-representable densities
may in principle be different.

Definitive work by Chayes {\em et al.} \cite{CCR85} proves that
on a grid, certain simple restrictions on the density
determine the set of ensemble $v$-representable densities (in both interacting and non-interacting cases).
This result explains why we were always able to find potentials for
a given density on a grid in 1d, where there is no degeneracy except for spin.
The work of Chayes {\em et al.} is reassuring, but not the final word on $v$-representability.
On a grid, the kinetic energy operator (proportional to the Laplacian) is always bounded,
whereas in the continuum it is not.
In such cases, inverting a density for the KS potential as in \Eqref{eqn:vSW} may lead to unacceptable
divergences, even for reasonable densities.
Proofs of $v$-representability on a grid \cite{K83} therefore do not guarantee
$v$-representability in the continuum.  
Complicating matters, properties which make for reasonable densities and potentials 
differ based on the dimensionality
of the problem \cite{L83}.
In this section, we therefore move away from 
our 1d grids and instead concentrate on
real 3d systems in the continuum.

In principle, one can invert any density $\n(\br)$ with $N\le 2$ for its KS potential
$v\s[\n](\br)$, as in \Eqref{eqn:vSW}.  Such an inversion, however, may lead to a potential 
which is singular and which does not have a well-defined ground state.
In order to avoid these problems, the potential should satisfy two key properties: 
(1) the KS Hamiltonian \eqref{eqn:KSeqn} being
bounded from below, and (2) the KS Hamiltonian being self adjoint \footnote{%
Self-adjointness can be
proven for certain sets of potentials.
For example, if the KS potential is in the $L^2 + L^\infty$ space, the
KS Hamiltonian is self-adjoint \cite{RSvol2}.
}.\nocite{RSvol2}
Properties which make the potential reasonable translate into properties
that the density should satisfy.  
In three dimensions, our reasonable potentials
are in the set $L^{3/2}+L^\infty$, 
which describes potentials of atoms, molecules, and solids \footnote{%
One may argue that there are many reasonable 3d potentials not in $L^{3/2}+L^\infty$,
such as the 3D isotropic harmonic oscillator:  $v(\br) = \half k r^2$.  Such potentials do not describe
real atoms, molecules, and solids, however, so we consider them unrealistic.
For a more tolerant viewpoint and in-depth mathematical discussion, see \Ref{L14}.}\nocite{L14}.
The density space whose dual is $L^{3/2}+L^\infty$ 
is $L^1 \cap L^3$, and this space is
a good start for the set of reasonable densities \cite{L83}. 
The $L^p$ space consists
of functions whose $p$ norm is finite:
\ben
L^p \equiv \left \{ f(\br) \ : \ \Big( \intr |f(\br)|^p \Big)^{1/p} < \infty \right \},
\een
where the integral is taken in the Lebesgue sense \cite{RSvol1}.
Thus our densities $\n(\br)$ should at least be in $L^1 \cap L^3$, and our
potentials in $L^{3/2} + L^\infty$. (This set includes Coulomb potentials \cite{L83}.) 
For a density whose inverted potential is
{\em not} in $L^{3/2} + L^\infty$ we say this density is non-$v$-representable.

To avoid unphysical densities, one 
should impose non-negativity and finite kinetic energy 
on the density, as articulated first by Lieb \cite{L83,CCR85,Leeuwen}: 
\ben
\intr n(\br) = N < \infty, \ \ n(\br)\geq 0 \ \ \forall\ \br ,\quad T\s\vW[n]<\infty,\ \ \label{eqn:restrictN}
\een
where the von Weizs\"acker kinetic energy is
\ben
T\s\vW[n] \equiv \intr \dfrac{|\nabla n(\br) |^2}{8\, n(\br) }, \label{eqn:vW}
\een
which is a lower-bound to the true kinetic energy $T[n]$ of the system.
We refer to such Lieb-allowed densities (which satisfy \Eqref{eqn:restrictN}) as reasonable.
Reasonable densities comprise a subset of $L^1 \cap L^3$ (by Sobolev's inequality, \Ref{L83}), 
so they have many useful properties. 
For example, for a reasonable density $n(\br)$ in a reasonable potential $v(\br)$
(i.e.\ $v(\br)$ is in $L^\infty+L^{3/2}$), the potential energy $|V[n]|<\infty$ \cite{L83}.
A density $n(\br)$ which fails to satisfy \Eqref{eqn:restrictN} can safely be regarded
as having an infinite $F[n]$ (or $T\s[n]$ for non-interacting systems) \cite{L83},  
and thus will be avoided 
in any iteration of the Kohn--Sham equations.
Reasonable densities are not always $v$-representable, however:
the inverted potential may not be in $L^\infty + L^{3/2}$.
But in these instances, there always exists a $v$-representable density $\tilde n(\br)$
that approximates the reasonable density $n(\br)$ to any desired accuracy, and which
allows the energies $F[\tilde n]$ and $T\s[\tilde n]$ to be calculated \cite{L83,Leeuwen}. 
In the remainder of this section, we will explore such an example
within the realm of non-interacting $v$-representability, or
$v\s$-representability for short.

We consider a density
which satisfies \Eqref{eqn:restrictN} but which is not $v\s$-representable.
Inspired by the fourth example of Englisch and Englisch \cite{EE83}, we choose:
\ben
n\PEE(\br) = A \big( 1 + | r - 1 |^{3/4} \big)^2 e^{-2r},
\label{eqn:baddens}
\een
where we normalize to two electrons with
\begin{eqnarray}
A&=&\frac{256e^2}{\pi\left(596e^2+273\, B+506\, C\right)}\\
&\approx&0.196521
\end{eqnarray}
with
\begin{eqnarray}
B&=&\sqrt{2\pi}\left(1+\frac2{\sqrt\pi}\int_0^{\sqrt2}dt\exp(t^2)\right)\\
C&=&\frac3{2^{\frac34}}\left(\Gamma\left[\frac34\right]-\int_0^2dt\frac{\exp(t)}{t^{\frac14}}\right).
\end{eqnarray}
This pathological density $n\PEE(\br)$ is not $v\s$-representable
due to the kink encountered at $r=1$,
which would require an inadmissible infinite-discontinuity in the KS potential.
To see this, we attempt to invert $n\PEE(\br)$ for its KS potential via \Eqref{eqn:vSW}:
\begin{eqnarray}
v\s[n\PEE](\br)&\overset{?}{=}&\frac12-\frac1r+\frac3{4\left(1+|r-1|^{3/4}\right)}\left[-\frac1{8|r-1|^{5/4}}\right.\nonumber\\
&&\left.+\frac{\delta(r-1)}{|r-1|^{1/4}}-\frac{\mathrm{sgn}(r-1)}{|r-1|^{1/4}}\left(1-\frac1r\right)\right],\label{eq:gedankenvs}
\end{eqnarray}
where we have used $\partial_x|x|=\mathrm{sgn}(x)$, $\partial_x^2|x|=2\delta(x)$, and 
$\mathrm{sgn}(x)$ is the sign function.   The worst offender is the
term proportional to $\delta(r - 1) / |r - 1|^{1/4}$ which fails to be in the set $L^{3/2}+L^\infty$. 
This makes $n\PEE(\br)$ non-$v\s$-representable, since it may not even be the ground state of
\Eqref{eq:gedankenvs}.  Furthermore, calculating $T\s[n]$ using the second-derivative formula of \Eqref{eqn:Ts}
is ill-defined, due to this discontinuity.
Nevertheless, $n\PEE(\br)$ is reasonable:
its $T\s\vW[n\PEE]$ is finite, as we will soon show.
So, despite the density being reasonable, 
it is non-$v\s$-representable. 
And while we are focusing on non-interacting  electrons, it is clear that 
$n\PEE(\br)$ would be troublesome for interacting electrons as well.

We obtain $T\s\vW[n\PEE]$ by first calculating its
kinetic energy density.  Due to spherical symmetry, we have:
\begin{eqnarray}
&&t\s\vW[n\PEE](r)=\half \left(\frac d{dr}\sqrt{n\PEE(r)}\right)^2\\
&& \ \ = \frac A 2 \left(  -1-|r-1|^{3/4}
+\frac{3\,\mathrm{sgn}(r-1)}{4\,|r-1|^{1/4}} \right)^2e^{-2r}, \ \ 
\end{eqnarray}
so that
\begin{eqnarray}
T\s\vW[n\PEE]&=&4\pi\int_0^\infty\!\! dr\,r^2\, t\s\vW[n\PEE](r)\\
&=&\frac{A\pi}{128e^2}\left(40e^2+93\, B-13\, C\right)\\
&\approx&0.996519.
\end{eqnarray}
Calculating $T\s[\n\PEE]$ via the second-derivative formula \eqref{eqn:Ts} 
seems like a simple integration by parts:
\ben
\left.
\begin{array}{rcl}
T\s[n] &=& \displaystyle -\half \intr \sqrt{n(\br)} \nabla^2 \sqrt{n(\br)}\\[10pt]
&=& - \displaystyle\intr n(\br)\, v\s[n](\br)
\end{array}
\right\} \quad(N\le 2),\label{TsN2}
\een
but due to the discontinuities
in $v\s[n\PEE](\br)$ \eqref{eq:gedankenvs}, this integral is ill-defined for $n\PEE(\br)$.

\begin{figure}
\includegraphics[width=\columnwidth]{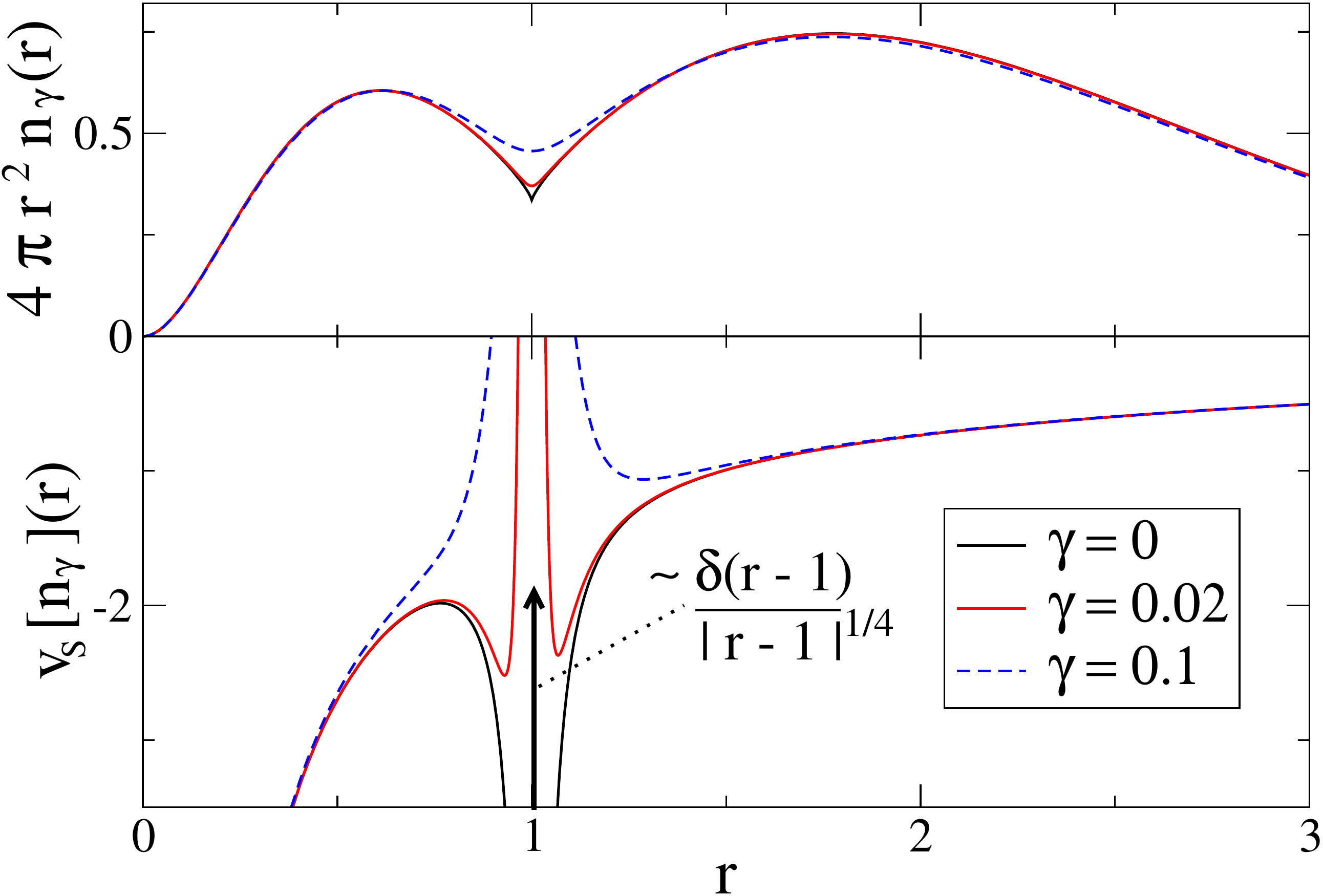}
\caption{
Inverting $n_\gamma(r)$ of \Eqref{smootheddens}
for the KS potential.  As $\gamma$ becomes smaller and smaller,
the changes in potential near $r=1$ become larger and larger.
}
\label{EEdens}
\end{figure}

We now illustrate how to obtain a $v\s$-representable density that is arbitrarily close
to our reasonable density $n\PEE(\br)$.  As a bonus, this procedure will also 
give a well-defined kinetic energy using the second-derivative formula.
Consider a function $f_\gamma(x)$ that
smooths out the $|r-1|$ in \Eqref{eqn:baddens}, but which has a parameter which can be 
continuously adjusted so that 
$\lim_{\gamma \to 0} f_\gamma(r-1) = |r-1|$.  We choose
\ben
f_\gamma(x) = \sqrt{x^2+\gamma^2},
\een
setting
\ben
n_\gamma(\br) = A_\gamma \big(1 + f^{3/4}_\gamma(r-1) \big)^2 e^{-2 r}.\label{smootheddens}
\een
(Note that the density must be renormalized for each value of $\gamma$.)
For small $\gamma$, the metric distance between $n\PEE(\br)$ and $n_\gamma(\br)$,
$\eta[n\PEE,n_\gamma]$ \eqref{etametric}, is proportional to $\gamma^{2.5}$; 
and $n_\gamma(\br)$ remains 
$v$-representable for all $\gamma > 0$.
In the iterations of the Kohn--Sham scheme, tolerances between densities are already
built into the method---namely as in \Eqref{eqn:tolerance}---so we need no
greater accuracy than that when finding a $v$-representable density close
enough to the target density.

As already mentioned, even though $T\s\vW[n\PEE]$ is finite,
$T\s[n\PEE]$ via \Eqref{TsN2} is ill-defined.
But by using the smoothed density of \Eqref{smootheddens}, we can calculate
$T\s[n_\gamma]$ and take the limit $\gamma \to 0$ (see \Figref{EETs}).  The result is the the same
as $T\s\vW[n\PEE]$, and this must be so based on simple mathematical considerations \cite{CCR85}. 
Two conjectures might be made after consider the foregoing:

\begin{figure}
\includegraphics[width=\columnwidth]{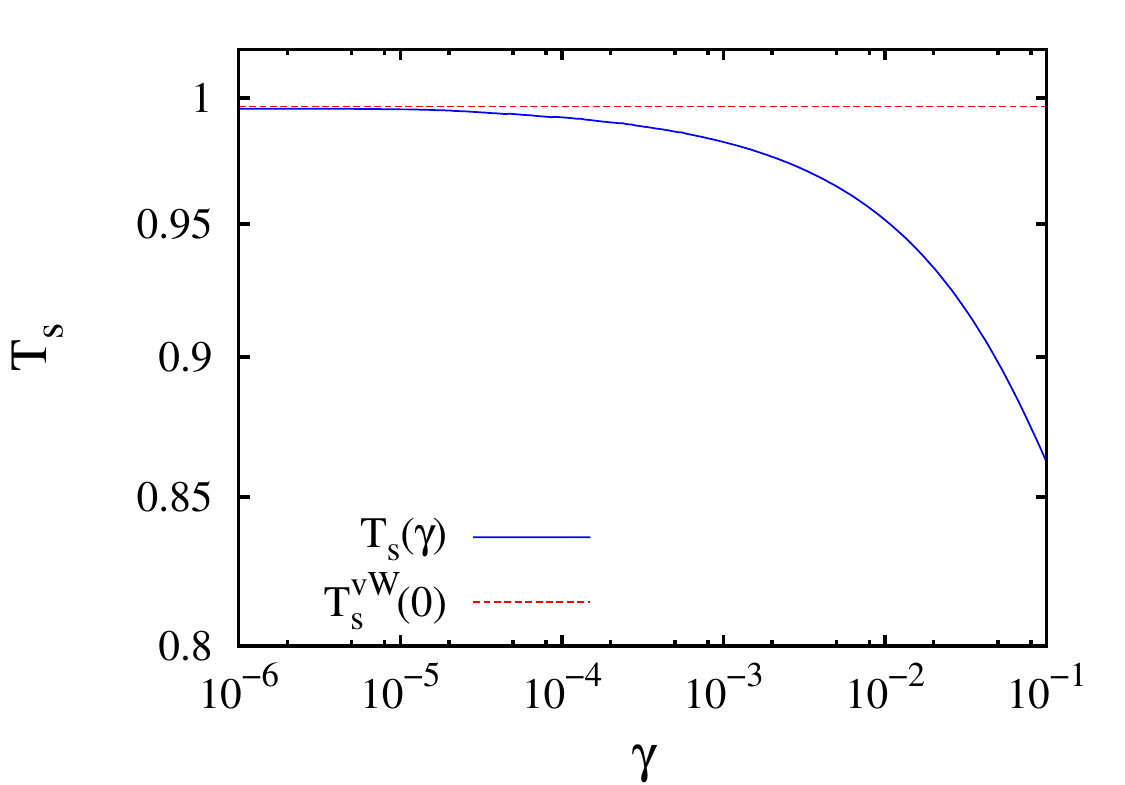}
\caption{ 
Kinetic energy convergence of $n\PEE(r)$ from Eq.~\eqref{smootheddens} 
by smoothing out 
the kink as in Eq.~\eqref{eqn:baddens}. While the von Weizs\"acker kinetic energy 
(vW) may be evaluated and integrated using only one derivative of 
the density, higher order derivatives of the density develop 
nonintegrable features. 
%
}
\label{EETs}
\end{figure}

\benu
\item A density being $v$-representable requires some bounds on the 
Laplacian (or second derivative) of the density.
On a grid, this is not an issue because
the Laplacian is always bounded.
\item Finite energies $F[n]$ and $T\s[n]$ 
may be extracted from reasonable but non-$v$-representable densities.
This can be done by suitably smoothing (or discretizing) the density and carefully
taking limits,
so as to remove divergent  terms.  For $N\le 2$, $T\s\vW[n]$ should give the
limit of $T\s[n]$ properly, and for $N > 2$ one should be able to use
\ben
T\s[n] = \sum_{j=1}^{N/2} \intr |\nabla \phi_j(\br)|^2,
\een
to avoid any singular divergences from second derivatives.
\enu

For some concluding remarks,
recall that the exact $E\Hxc[n]$ is defined using both interacting and non-interacting
systems.  This means that we need $n(\br)$ to be $v\s$ and $v$-representable
to calculate $E\Hxc[n]$.  While in principle 
$v\s$-representable densities comprise a different set than $v$-representable densities,  
we can use the methods of this section to calculate $E\Hxc[\n]$ for any reasonable density.
The prescription is to find a $v\s$-representable density $\tilde n\s(\br)$ and a $v$-representable
density $\tilde n(\br)$ which are within some small tolerance of $n(\br)$ and each other.
With the inverted potentials $\tilde v\s(\br)$ and $\tilde v(\br)$, self-consistent KS calculations are possible,
given $v\Hxc[\n](\br) = \tilde v\s(\br) - \tilde v(\br)$ as in \Eqref{eqn:vHxc}.
We hope to further explore the connections between interacting and 
non-interacting $v$-representability in future work.

As a final note, all of our numerical inversions have used pure-state wavefunctions.
This is justified for spin-singlet 1d systems and for this simple spin-singlet example in 3d.
In systems with degeneracy, however, the ensemble formulation of DFT
should be used, not only because the ensemble
$E_v[n]$ is convex \cite{WSBW13}, but also because 
the class of pure-state $v$-representable densities is smaller than the class 
of ensemble $v$-representable densities \cite{L83,Leeuwen,SGB97}. 
Outside of this section, we always worked on a grid, which means that
$v$-representability difficulties were not an issue \cite{CCR85}.  We 
found no cases where, as the grid spacing goes to zero, the potential diverged as in 
the example here.

\sec{Conclusions}

Our investigations into the exact functional demonstrate that it is possible to
solve the Kohn--Sham equations with the exact XC functional 
for simple model systems at great computational cost.
Our calculations involve mapping  the functional landscape for more than
just the ground-state density, enabling us to address questions of convergence
within the KS scheme.  We tested many systems, and found that  
strongly correlated systems pose a greater challenge,
not only from a theoretical standpoint in finding accurate approximations, 
but also practically within the KS scheme,
where smaller steps must be taken (or more sophisticated methods used) 
to converge the calculation.  In a word, the exact functional landscape 
for strongly correlated systems is more treacherous, but not impossible, for
a simple  KS algorithm to navigate.

Despite the surmountable convergence difficulties for strongly correlated systems, 
the only stationary point of the KS equations
is the ground-state density of the original problem, given $v$-representable densities as inputs.
This is simply a reaffirmation
of the HK theorem, that there is a one-to-one correspondence between ground state densities
and potentials.  This is the case even for stretched systems, where approximate
functionals would prefer to break spin symmetry; the exact spin-density functional
has only one stationary point, at the correct ground-state spin densities.  All changes
in density away from that point cause the energy to rise.  Thus the lowest energy 
stationary point with an approximate functional has the same energy landscape as the
true functional, and should be treated as the prediction for the energy with that approximation,
regardless of how many symmetries have been broken.  This reaffirms the conclusions of \Ref{PSB95}.

The density mixing algorithm used to prove convergence of the KS scheme
is one of the simplest ways to explore
the infinite-dimensional set of possible densities, and it provides insight into the gradient-descent
nature of the KS scheme.
While this algorithm is too primitive for modern practical implementations, its
main purpose here is to provide a definite framework in which convergence questions
can be studied.

There is another avenue of research, but which cannot be pursued
in these model 1d systems: the effects of orbital degeneracy
within exact KS theory, especially due to angular momentum.
An ensemble of degenerate densities may easily 
not be pure-state $v$-representable \cite{L83,Leeuwen}, and the extent 
of the challenges for exact DFT warrants investigation.
Unfortunately, this avenue cannot be explored for these 1d systems, in which there
is no angular momentum.  Exploring these concepts in 3d would shed light
on how DFT handles strong correlation effects due to {\em exact} degeneracies, in contrast 
to the near degeneracies \cite{HG11} we have investigated (e.g.\ in stretched H$_2$) 
for which exact DFT performs well in 1d \cite{WSBW13}.

Finally, we discuss the consequences of our example of a non $v$-representable density.
The example we give is a reasonable density, meaning it is in the domain on with the
Levy--Lieb density functional is defined:  it is normalized, non-negative, and has
finite kinetic energy.   Consistent with the proof of Chayes et.~al.~\cite{CCR85}, on any finite
grid, it has a well-behaved Kohn--Sham potential.  But as the grid-spacing is brought
to zero, divergences appear in that potential, so that it is ill-defined in the
continuum limit.  So this is an example of a density that is $v$-representable on
a lattice, but is not $v$-representable in the continuum.  Similarly, one can remain
in the continuum and introduce a small parameter ($\gamma$) which rounds off the cusp
in the density.  For any finite value of $\gamma$, no matter how small, the potential
is finite and well-behaved.   Thus our cuspy density is arbitrarily close to a $v$-representable
density.  These are the standard arguments given in the physics literature for why $v$-representability
is not an issue in DFT.

But our example shows that there is still something to worry about.  Either regularization
procedure (finite grid spacing or finite $\gamma$) fails in the limit, and anyone doing
an inversion on such a density should check their KS potential converges to a well-defined
limit.   Our example density fails this test.

The important question is not whether some artificially created density is $v$-representable
or not.  The real question is, given the densities of atoms, molecules, and solids, i.e.,
densities generated by solving the Schr\"odinger equation with Coulomb interactions, are there
features like that of our example that produce ill-behaved KS potentials?   This is all
that matters, and practical experience suggests that such situations are rare, if they occur
at all.

The authors would like to gratefully acknowledge the support of
DOE grant DE-SC0008696  for funding this work.
LW would also like to thank
the Korean government for additional funding 
through the global research network grant (No.~NRF-2010-220-C00017).

\bibliography{ksdft}

\end{document}